%% file: main.tex
\documentclass[journal=jacsat,manuscript=article]{achemso}

\usepackage{amsmath}
\usepackage[utf8]{inputenc} 
\usepackage[T1]{fontenc}    
\usepackage{hyperref}       
\usepackage{url}            
\usepackage{booktabs}       
\usepackage{amsfonts}       
\usepackage{nicefrac}       
\usepackage{microtype}      
\usepackage{cleveref}       
\usepackage{lipsum}         
\usepackage{graphicx}
\usepackage{natbib}
\usepackage{doi}
\usepackage{placeins}
\usepackage{amssymb}
\usepackage{multirow}
\usepackage{siunitx}
\usepackage{tabularx}
\usepackage{tocloft}

\usepackage{hyperref}

\AtBeginDocument{}

\author{Carlos Bornes}
\affiliation{Department of Physical and Macromolecular Chemistry, Charles University, Hlavova 8, Praha 2, Prague 12800, Czech Republic}
\author{Chiheb Ben Mahmoud}
\affiliation{Department of Materials, University of Oxford, Oxford OX1 3PH, United Kingdom}
\author{Volker L. Deringer}
\affiliation{Inorganic Chemistry Laboratory, Department of Chemistry, University of Oxford, Oxford OX1 3QR, United Kingdom}
\author{Christopher J. Heard}
\affiliation{Department of Physical and Macromolecular Chemistry, Charles University, Hlavova 8, Praha 2, Prague 12800, Czech Republic}
\author{Lukáš Grajciar}
\affiliation{Department of Physical and Macromolecular Chemistry, Charles University, Hlavova 8, Praha 2, Prague 12800, Czech Republic}
\email{grajciar@natur.cuni.cz}

\title[]{An Accurate Tensorial Model for Prediction of Full Zeolite NMR Spectra}

\begin{document}
\maketitle
\begin{abstract}
Solid state nuclear magnetic resonance (ss-NMR) is one of the most sensitive and popular techniques for structure elucidation in geometrically complex crystalline materials, such as zeolites. Synergistic support from computational modelling is vital to interpret experimental spectra, and relate ss-NMR to atomistic models. Nevertheless, computational predictions are hindered by the high expense of calculating magnetic shielding (MS) and electric field gradient (EFG) tensors from first principles. In this work, we leverage a novel tensorial machine learning approach to train a general model for predicting complete NMR tensors. We demonstrate the utility of the approach for a diverse dataset of zeolitic materials and NMR-active nuclei ($^{27}$Al, $^{29}$Si, $^{17}$O, $^{23}$Na and $^{1}$H), predicting all NMR observables to a high degree of precision. These observables are then translated into predictions of the full $^{27}$Al and $^{29}$Si ss-nMR spectra for the exemplary zeolite RTH. Thus, this work opens a pathway to accurate, high-throughput NMR simulation for large-scale and realistic models of chemically complex zeolites.
\end{abstract}


\section{Introduction}
Zeolites are crystalline nanoporous aluminosilicates widely used in heterogeneous catalysis, gas separation, and water purification.\cite{flanigen2010Introduction, roth2015Zeolites21sta} Despite decades of academic research and industrial deployment, many fundamental long-standing questions remain, including the preference of specific aluminum distributions and speciation, the effects of cation solvation and dynamics, and the structure and formation mechanism of defects. One reason hindering the advance of a true atomistic understanding of zeolites arises from the disconnect between complex zeolite samples used experimentally and the simplified models typically employed in DFT-based studies. 

Among experimental techniques capable of bridging this gap, solid-state nuclear magnetic resonance (NMR) spectroscopy stands out as one of the most powerful techniques used for probing zeolite local structure and dynamics,\cite{polenova2015MagicAngle, mafra2012StructuralCharacterization} with nuclei such as $^{1}$H, $^{17}$O, $^{23}$Na, $^{27}$Al, and $^{29}$Si each providing a unique window into different aspects of zeolite chemistry. However, the interpretation of NMR spectra is often hampered by the complexity of real samples, where multiple overlapping environments, the effects of hydration and temperature, and disorder make unambiguous spectral assignment challenging.\cite{lei2023needoperandoa} First-principles calculations of NMR parameters via the gauge-including projector augmented wave (GIPAW) method\cite{pickard2001Allelectronmagnetic,yates2007CalculationNMR} have become essential tools for bridging this gap,\cite{bonhomme2012FirstPrinciplesCalculation,ashbrook2016Combiningsolidstate} but their high computational cost restricts applications to small unit cells, static or short \textit{ab initio} molecular dynamics (MD) trajectories, and a limited subset of possible configurations.

Machine learning interatomic potentials (MLIPs) have partially addressed the computational bottleneck by enabling large-scale atomistic simulations that retain the accuracy of \textit{ab initio} methods at a fraction of the cost, enabling the exploration of large configurational spaces and long timescales relevant to experimental conditions. MLIPs now span a broad range of specificity, from reaction-specific potentials,\cite{bocus2025OperandoNature, achar2025reactive} through class-specific models targeting families of materials such as silicates, ranging from the Si--O system,\cite{erhard2024modelling} SiO$_2$--water interfaces,\cite{roy2024learning} and hydrated aluminosilicate zeolites,\cite{erlebach2022Accuratelargescalea,bocus2023Nuclearquantum,erlebach2024reactiveneurala, brugnoli2024neural} to foundational models trained on diverse datasets,\cite{batatia2025foundationmodel,neumann2024OrbFast,mazitov2025PETMADlightweight} each offering different trade-offs between accuracy and transferability. Despite these advances, MLIPs only predict energies and forces for driving MD simulations, while the prediction of experimentally observable properties such as NMR parameters often requires a separate ML model and remains comparatively unexplored, owing in part to the higher cost of generating reference datasets and the need for ML architectures capable of learning tensorial response properties that transform equivariantly under rotations.

For molecular solids, the ShiftML framework\cite{paruzzo2018Chemicalshifts} has demonstrated the power of SOAP-based kernels for predicting chemical shifts of $^{1}$H, $^{13}$C, $^{15}$N, and $^{17}$O, and in its most recent iterations the ShiftML was expanded to other nuclei,\cite{cordova2022MachineLearning} and the prediction of chemical shift anisotropy.\cite{kellner2025deeplearning} However, ShiftML is trained on organic molecular solids and does not cover nuclei central to zeolite chemistry such as $^{23}$Na, $^{27}$Al, or $^{29}$Si, nor does it predict electric field gradient (EFG) tensors. The latter is critical from materials science as over 75\% of all NMR-active nuclei are quadrupolar.\cite{ashbrook2009Recentadvances} 

More broadly, early studies using ML for predicting NMR properties focused on predicting scalar quantities (isotropic chemical shifts), and beyond those, it seems timely to make use of the rich structural and dynamical information encoded in tensorial NMR parameters such as chemical shift anisotropy (CSA) and quadrupolar coupling parameters ($C_Q$, $\eta_Q$).\cite{ashbrook2016Combiningsolidstate,eden2023Probingoxidebased} The prediction of these tensorial observables poses a fundamental challenge because, unlike scalar properties, they are not rotationally invariant. Recent progress has been made using rotationally equivariant descriptors such as $\lambda$-SOAP\cite{charpentier2025FirstprinciplesNMR,harper2025Performancemetrics} and equivariant graph neural networks\cite{benmahmoud2025Graphneuralnetworkpredictions,venetos2023MachineLearning} for predicting NMR observables. In particular, Ben Mahmoud \textit{et al.}\cite{benmahmoud2025Graphneuralnetworkpredictions} demonstrated that equivariant graph neural networks (eGNN), using the NequIP \cite{batzner2022E3equivariantgraph} MLIP architecture at the time, can predict full NMR magnetic shielding and EFG tensors for amorphous SiO$_2$ with high accuracy. However, whether such approaches can achieve similar accuracy across chemically and structurally diverse material classes, such as aluminosilicate zeolites with varying Si/Al ratios, multiple cation types, and different hydration states has not been demonstrated. 

In the specific context of zeolites, early works exploited simple correlations between $^{29}$Si and $^{27}$Al isotropic chemical shifts and local geometric parameters such as $T$--$O$--$T$ angles.\cite{ramdas1984simplecorrelationa, lippmaa1986Highresolutionaluminum27} Over the years, the method has been used and expanded to predict mainly the $^{29}$Si NMR of siliceous zeolites. Recently, we have employed least absolute shrinkage and selection operator (LASSO) regression to rapidly predict $^{27}$Al and $^{23}$Na chemical shifts in aluminosilicate zeolites,\cite{lei2023needoperandoa,lei2025machinelearningb} which retains structural interpretability but is constrained to low dimensionality and linearity, and requires the user to choose the relevant features. Smooth Overlap of Atomic Positions (SOAP) descriptors combined with non-linear regression methods, such as kernel ridge regression, have been found to sacrifice interpretability but achieve higher accuracy for isotropic shifts across diverse zeolite frameworks.\cite{willimetz202527NMRa,willimetz2025AluminumSiting,chaker2019NMRshiftsa} 

In this work, we develop a graph-based tensorial model for NMR parameters in aluminosilicate zeolites, capable of predicting complete NMR magnetic shielding and electric field gradient tensors for five nuclei ($^{1}$H, $^{17}$O, $^{23}$Na, $^{27}$Al, and $^{29}$Si) across the  compositional and structural diversity of this highly complex material class. Our model is trained on a diverse dataset spanning pure silica and aluminum-containing zeolites with different Si/Al ratios, protonic and sodium charge-compensating cations, varying water loadings, and multiple framework topologies. We demonstrate that this model accurately reproduces DFT-calculated NMR parameters and we validate its predictions against experimental data from the literature.\cite{willimetz2025AluminumSiting}
We also discuss more general implications for the development of tensor NMR models for a broader range of inorganic materials.

\section{Materials and methods}
\subsection{Data generation}
To ensure broad coverage of the chemical and configurational space of aluminosilicate zeolites, we used a subset of structures from previous DFT databases developed for training MLIPs in hydrogen-\cite{erlebach2024reactiveneurala} and sodium-containing\cite{lei2025machinelearningb} zeolites. These databases encompass a wide range of zeolite frameworks, both existing and hypothetical, Si/Al ratios, charge-compensating cations (H$^+$ and Na$^+$), and water loadings. The original datasets were generated using \textit{ab initio} MD simulations at multiple temperatures (1200--3600~K) to sample reactive events, complemented by systematic lattice deformations to cover low-energy regions of the potential energy surface. The databases also include dense aluminosilicates, alumina, and silica polymorphs, bulk water at various densities and temperatures, water clusters \textit{in vacuo}, and ice polymorphs to ensure transferability across different structural motifs and environmental conditions. For detailed information on the database construction, including specific framework selection, MD protocols, and lattice deformation strategies, we refer the reader to our previous works.\cite{erlebach2024reactiveneurala,lei2025machinelearningb}

The combined databases encompassing around 300 000 structures of H-/Na-exchanged (alumino)silicates was subsampled to extract a structurally diverse subset suitable for NMR calculations. Structures were selected using the farthest point sampling (FPS) algorithm applied to MACE representations to maximize structural diversity in the descriptor space.\cite{batatia2022MACE} This procedure yielded around 12 000 configurations spanning the full range of chemical compositions, framework topologies, and environmental conditions present in the original databases. NMR magnetic shielding and electric field gradient (EFG) tensors were calculated for all selected structures at the DFT level using the gauge-including projector augmented wave (GIPAW) method\cite{pickard2001Allelectronmagnetic,yates2007CalculationNMR} as implemented in CASTEP.\cite{clark2005Firstprinciples} Calculations employed on-the-fly generated ultrasoft pseudopotentials, the PBEsol exchange-correlation functional, a $k$-point spacing of 0.05~\AA$^{-1}$, and a plane-wave energy cutoff of 900~eV. To ensure the proper convergence with regard to relevant NMR tensor observables, such as the isotropic chemical shift ($\delta_\mathrm{iso}$), chemical shift anisotropy parameters span ($\Omega$) and skew ($\kappa$), and electric field gradient parameters, quadrupolar coupling constant ($C_Q$) and asymmetry parameter ($\eta_Q$), we selected 10 structurally diverse structures from the entire database, including H- and Na-zeolites with different Al, a bulk Na-aluminosilicate, water, SiO$_2$ and Al$_2$O$_3$ polymorphs to conduct the convergence tests (Figures \ref{fig:dft-conv1}-\ref{fig:dft-conv10}).

Although all $\sim$12\,000 FPS-selected configurations produced valid GIPAW calculations, we found that training ML models directly on the full dataset led to significantly degraded prediction accuracy compared to a curated subset. This can be attributed to high-energy configurations sampled from high-temperature AIMD, which exhibit distorted geometries and extreme NMR parameters outside the range relevant to ground-state structures. This behavior contrasts with MLIP training for energies and forces, where all $\sim$12\,000 structures can be used without significant performance loss, and suggests that learning NMR tensor components is more sensitive to outlier configurations than learning scalar energy labels. To identify and remove problematic structures, we applied an interquartile range (IQR) outlier analysis to the $\sigma^{(0)}$ prediction errors of preliminary models trained on the complete dataset. Structures for which any nucleus exhibited an $\sigma^{(0)}$ prediction error exceeding $k \times \mathrm{IQR}$ above the third quartile or below the first quartile were removed. We systematically evaluated scale factors of $k = 2$, $3$, and $4$ (Tables~\ref{tab:iqr-l0-ms}--\ref{tab:iqr-l2-ms}). Progressively stricter filtering consistently improved model accuracy across all nuclei and all tensor components ($\sigma^{(0)}$, $\sigma^{(1)}$, and $\sigma^{(2)}$), with particularly dramatic reductions in RMSE---for example, the $^{17}$O $\sigma^{(0)}$ RMSE dropped from 164.8~ppm (full dataset) to 5.7~ppm ($k = 2$). The final dataset, obtained with $k = 2$, comprises 7027 structures, as this threshold yielded the lowest prediction errors across all nuclei and tensor ranks (Tables~\ref{tab:iqr-l0-ms}-\ref{tab:iqr-l2-ms}).

\subsection{NMR model training}
Ben Mahmoud \textit{et al.}\cite{benmahmoud2025Graphneuralnetworkpredictions} recently demonstrated that NequIP,\cite{batzner2022E3equivariantgraph, tan2025Highperformancetraining} an eGNN can be utilised to predict tensorial properties, including magnetic shielding and EFG tensors, by leveraging their internal spherical tensor representations. In this work, we employ the TensorMACE\cite{benmahmoud2026TensorMACE} model implemented in Graph-PES,\cite{gardner2024graphpestrain} using the MACE architecture\cite{batatia2022MACE} to predict tensorial properties. In contrast to NequIP, which directly learns irreducible spherical tensors, TensorMACE reconstructs target spherical tensors through tensor products of internal equivariant features.\cite{gardner2024graphpestrain, benmahmoud2025Graphneuralnetworkpredictions}

All results presented in the following sections correspond to models trained on the IQR-filtered dataset ($k = 2$) containing 7027 structures, with an 80/10/10 split for training, validation, and testing. The impact of dataset curation on model accuracy is detailed in Tables~\ref{tab:iqr-l0-ms}--\ref{tab:iqr-l2-ms} for magnetic shielding and Table \ref{tab:iqr-l2-efg} for the EFG tensor. Optimization of hyperparameters showed that key hyperparameters include the angular resolution ($L_\mathrm{max}$), correlation order, cutoff radius, number of channels, and the number of tensor products used to reconstruct the target NMR tensors. All models were trained using a cutoff radius of 5.5~\AA, 3 interaction layers, a $L_\mathrm{max}$ = 3, and 256 channels per irreducible representation type per interaction layer.

\subsection*{Model benchmarking}
Molecular dynamics simulations were performed for 1 ns using our previously developed MLIPs\cite{erlebach2024reactiveneurala, lei2025machinelearningb} for siliceous RTH, and for Al-containing H-RTH with one aluminum atom placed at each of the four crystallographically distinct T sites, in both hydrated and dehydrated forms.\cite{willimetz2025AluminumSiting}  From each trajectory, 50 equally spaced snapshots were extracted, and NMR tensors were predicted using the ML model for each snapshot. For the Al-containing systems, trajectories with aluminum at different T sites were calculated separately for the hydrated and dehydrated forms. Time-averaged magnetic shielding and EFG tensors were computed for each crystallographic site by averaging over all snapshots, using the Soprano code.\cite{sturniolo2025Soprano} The resulting tensors were used to simulate $^{29}$Si and $^{27}$Al MAS NMR spectra at 400 and 750 MHz using SIMPSON,\cite{bak2011simpson, tovsner2014computer, juhl2020versatile} with the isotropic chemical shift, anisotropy, asymmetry, and principal axis orientation for each interaction extracted from the averaged tensors. DFT reference spectra were generated using the same protocol, with GIPAW-calculated tensors replacing ML predictions. SIMPSON input files were generated using Soprano and Simpyson.\cite{sturniolo2025Soprano, simpyson_v0.2}

\section{Results and discussion}
\subsection{Prediction of magnetic shielding tensors}
Rank-2 Cartesian tensors, such as NMR tensors, can be decomposed into irreducible spherical components of rank $\sigma^{(0)}$, $\sigma^{(1)}$, and $\sigma^{(2)}$, which transform as spherical harmonics under rotation.\cite{mueller2011Tensorsrotations,man2013CartesianSpherical} This decomposition is particularly well suited for eGNN, since each irreducible component has well-defined rotational transformation properties that can be directly encoded in the model.\cite{benmahmoud2025Graphneuralnetworkpredictions,grisafi2018SymmetryAdaptedMachine} For NMR tensors, the decomposition separates physically distinct contributions. The magnetic shielding tensor is typically an asymmetric $3 \times 3$ tensor, and its decomposition yields three irreducible components: a scalar of rank $\sigma^{(0)}$, corresponding to the isotropic magnetic shielding $\sigma_\mathrm{iso}$, a pseudovector of rank $\sigma^{(1)}$ with 3 independent components, representing the antisymmetric part of the tensor, and a symmetric traceless tensor of rank $\sigma^{(2)}$ with 5 independent components, which encodes the CSA, and observables such as $\Omega$ and $\kappa$.

\begin{figure}[h!]
  \centering
  \includegraphics[width=0.9\textwidth]{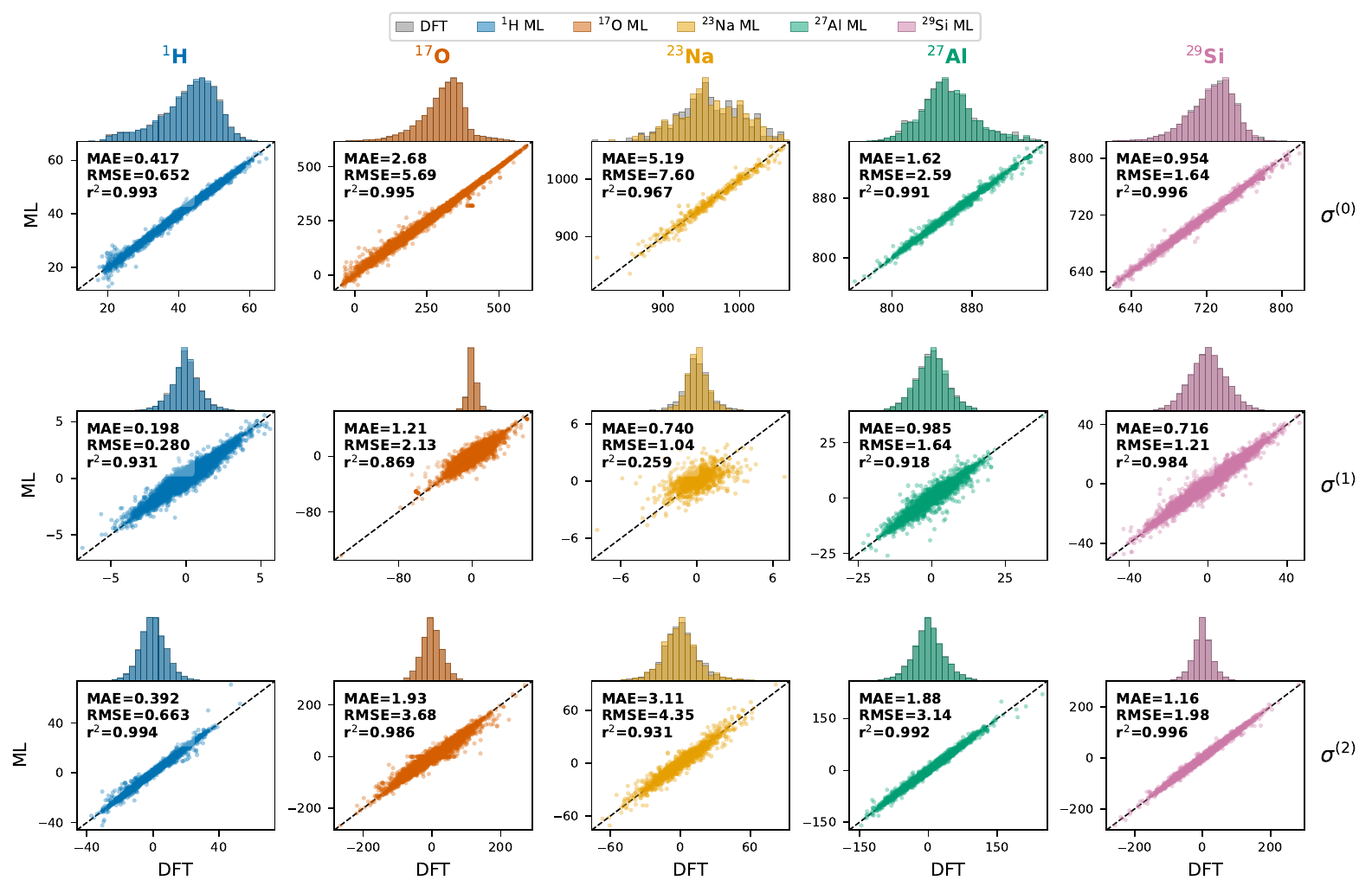}
  \caption{Parity plots comparing DFT-calculated and ML-predicted irreducible spherical tensor components ($\sigma^{(0)}$, $\sigma^{(1)}$, and $\sigma^{(2)}$) of the magnetic shielding tensor for $^{1}$H, $^{17}$O, $^{23}$Na, $^{27}$Al, and $^{29}$Si on the test set. Marginal histograms show the distribution of DFT (gray) and ML (colored) values. All values are in ppm.}
  \label{fig:test-parity-ms}
\end{figure}

Figure~\ref{fig:test-parity-ms} shows the parity plots for the magnetic shielding tensor components on the test set, comparing ML predictions against DFT reference values for all five nuclei ($^{1}$H, $^{17}$O, $^{23}$Na, $^{27}$Al, and $^{29}$Si). The isotropic component $\sigma^{(0)}$ is predicted with high accuracy across all nuclei, with MAEs of 0.42~ppm ($^{1}$H), 2.68~ppm ($^{17}$O), 5.19~ppm ($^{23}$Na), 1.62~ppm ($^{27}$Al), and 0.95~ppm ($^{29}$Si), and $R^2$ values exceeding 0.96 in all cases. The $\sigma^{(2)}$ components exhibit similarly strong predictive performance, with MAEs ranging from 0.39 to 3.11 and $R^2 > 0.93$ for all nuclei. Similar to earlier findings,\cite{benmahmoud2025Graphneuralnetworkpredictions} the antisymmetric $\sigma^{(2)}$ components are consistently the most challenging to predict. For $^{1}$H, $^{27}$Al, and $^{29}$Si, the model still achieves good accuracy ($R^2 = 0.93$, $0.92$, and $0.98$, respectively), while for $^{17}$O the performance is moderate ($R^2 = 0.87$). The most notable difficulty arises for $^{23}$Na, where the $\sigma^{(1)}$ component shows a low $R^2 = 0.26$, although the absolute errors remain small (MAE~=~0.74 and RMSE~=~1.04 ppm). This behavior can be understood from the narrow distribution of $\sigma^{(1)}$ values for sodium, in combination with the much smaller number of $^{23}$Na training points (Table~\ref{tab:num-atom-types}). From a practical standpoint, however, this limitation has minimal impact on experimentally relevant observables, since the antisymmetric shielding contribution is typically unresolvable in standard solid-state NMR experiments\cite{anet1990NMRrelaxation} and does not affect commonly measured parameters such as $\delta_\mathrm{iso}$, $\Omega$, or $\kappa$. Consistent performance across training, validation, and test sets (Figures~\ref{fig_si:train-parity-ms}--\ref{fig_si:valid-parity-ms}) indicates that the model generalizes well.

\subsection{Prediction of electric field gradient tensors}

In contrast to magnetic shielding tensors, a EFG tensor $V$ is by definition symmetric and traceless, and therefore contains only $V^{(2)}$ components, and thus the model should learn only the 5 irreducible spherical components. The conversion of the EFG tensor from Cartesian to irreducible spherical components as expected resulted in an $V^{(1)}$ vector of zeros, however, likely due to numerical noise we observed some non-zero $V^{(0)}$ values on the order of $10^{-11}$, which were removed for training.

In the last few years the prediction of EFG tensors, or tensor observables such as $C_Q$ and $\eta_Q$ has gained some traction,\cite{sun2025Machinelearning} but these efforts are typically limited to chemically and configurationally narrow datasets. Harper \textit{et al.}\cite{harper2025Performancemetrics} recently investigated performance metrics for assessing the learning of EFG tensors and demonstrated the superiority of a tensorial learning approach over separate scalar learning of individual tensor-derived observables. Charpentier\cite{charpentier2025FirstprinciplesNMR} used $\lambda$-SOAP descriptors combined with linear ridge regression to predict EFG components for $^{17}$O and $^{23}$Na in Na$_2$O--SiO$_2$ glasses with a high degree of accuracy. Ben Mahmoud \textit{et al.}\cite{benmahmoud2025Graphneuralnetworkpredictions} demonstrated that equivariant graph neural networks can predict full EFG tensors with high accuracy for $^{17}$O in SiO$_2$ polymorphs. Our work extends these approaches to a much broader chemical space, demonstrating accurate prediction of EFG tensors for five different nuclei across diverse aluminosilicate zeolites.

\begin{figure}[h!]
  \centering
  \includegraphics[width=0.9\textwidth]{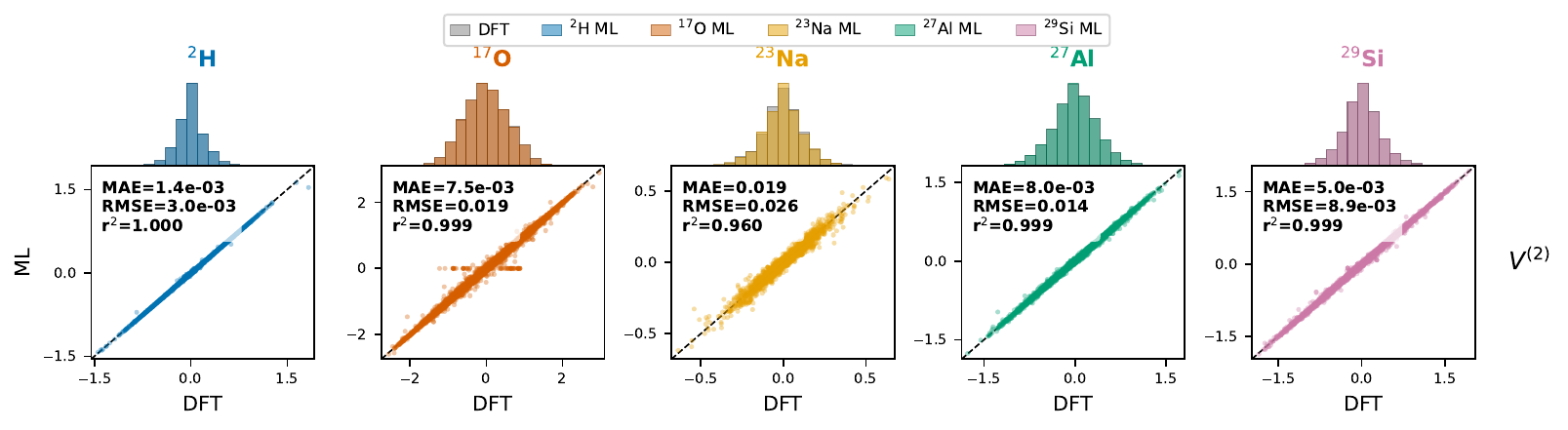}
  \caption{Parity plots comparing DFT-calculated and ML-predicted $V^{(2)}$ components of the electric field gradient tensor for $^{2}$H, $^{17}$O, $^{23}$Na, $^{27}$Al, and $^{29}$Si on the test set. The EFG tensor is symmetric and traceless and therefore contains only $V^{(2)}$ components. All values are in atomic units.}
  \label{fig:test-parity-efg}
\end{figure}

\FloatBarrier
The prediction of EFG tensors (Figure~\ref{fig:test-parity-efg}) shows very low MAEs for all nuclei: 0.001~a.u. ($^{2}$H), 0.007~a.u. ($^{17}$O), 0.019~a.u. ($^{23}$Na), 0.008~a.u. ($^{27}$Al), and 0.005~a.u. ($^{29}$Si). Since the magnetic shielding and EFG tensors have different units (ppm vs.\ a.u.) and value ranges, their absolute MAE values cannot be directly compared. To enable a fair cross-property comparison, we adopt the \%RMSE defined as the RMSE normalized by the standard deviation of the DFT reference values. For the shared $\sigma^{(2)}$ and $V^{(2)}$ components, the \%RMSE is consistently lower for EFG than for magnetic shielding across all nuclei (Table~ \ref{tab:iqr-l2-ms} and \ref{tab:iqr-l2-efg}), for example 12.03\% vs.\ 3.71\% for $^{17}$O and 8.76\% vs.\ 4.36\% for $^{27}$Al, for magnetic shielding and EFG tensors respectively. This confirms that EFG tensors are genuinely easier to learn and not merely smaller in absolute magnitude. One way to rationalise the better learning performance of the EFG tensor is that for the MS tensor, one is trying to simultaneously fit three qualitatively different irreducible components ($\sigma^{(0)}$, $\sigma^{(1)}$, and $\sigma^{(2)}$) with competing loss contributions, whereas the EFG model learns only $V^{(2)}$ components. A similar performance is observed across training and validation sets (Figures~\ref{fig_si:train-parity-efg}--\ref{fig_si:valid-parity-efg}).

\FloatBarrier
\subsection{Prediction of tensor observables}
Having established the accuracy of the model for predicting the irreducible spherical tensor components, we now assess its performance for the experimentally relevant tensor observables that are routinely extracted from NMR spectra. For magnetic shielding tensors, these are the isotropic magnetic shielding $\sigma_\mathrm{iso} = \mathrm{Tr}(\boldsymbol{\sigma})/3$, the span $\Omega = \sigma_{11} - \sigma_{33}$, and the skew $\kappa = 3(\sigma_{22} - \sigma_\mathrm{iso})/\Omega$; for EFG tensors, the quadrupolar coupling constant $C_Q = eQV_{zz}/h$ and the asymmetry parameter $\eta_Q =(V_{xx}-V_{yy})/V_{zz}$, , where $e$ is the elementary charge, $Q$ the nuclear quadrupole moment, $h$ Planck's constant, and $V_{ii}$ the eigenvalues of the EFG tensor ordered so that $|V_{zz}| \geq |V_{yy}| \geq |V_{xx}|$.

\begin{figure}[h!]
  \centering
  \includegraphics[width=0.9\textwidth]{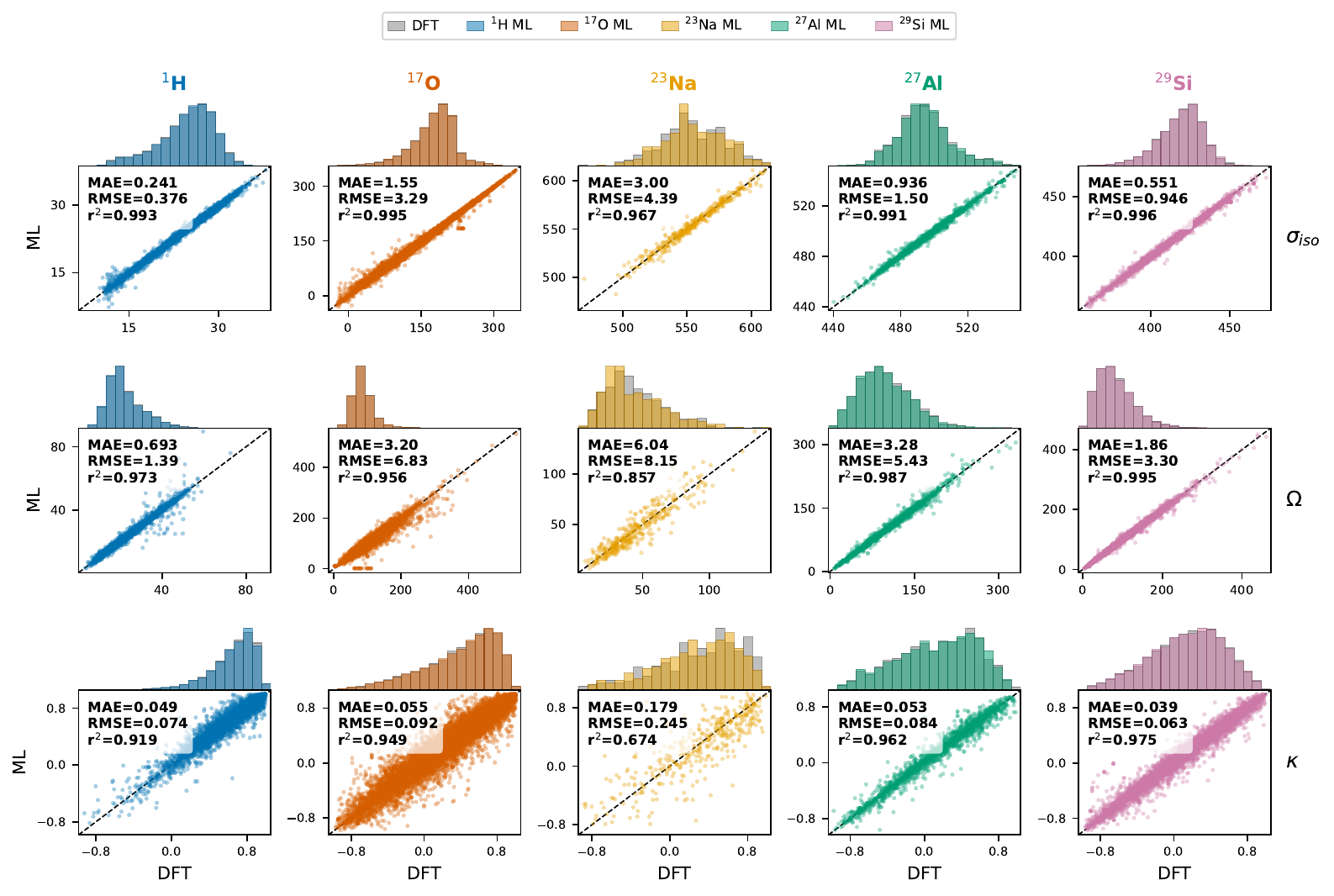}
  \caption{Parity plots comparing DFT-calculated and ML-predicted magnetic shielding tensor observables for $^{1}$H, $^{17}$O, $^{23}$Na, $^{27}$Al, and $^{29}$Si on the test set. The isotropic chemical shift ($\delta_\mathrm{iso}$), span ($\Omega$), and skew ($\kappa$) are shown. All values are in ppm.}
  \label{fig:test-parity-ms-observables}
\end{figure}

Figure~\ref{fig:test-parity-ms-observables} shows the parity plots for the magnetic shielding observables on the test set. The isotropic shielding $\sigma_\mathrm{iso}$ is predicted with excellent accuracy for all five nuclei, with MAEs of 0.24~ppm ($^{1}$H), 1.55~ppm ($^{17}$O), 3.00~ppm ($^{23}$Na), 0.94~ppm ($^{27}$Al), and 0.55~ppm ($^{29}$Si), and $R^2 \geq 0.97$ in all cases. Note that the lower MAE for $\sigma_\mathrm{iso}$ relative to the $\sigma^{(0)}$ spherical component is a consequence of the normalization convention $\sigma_\mathrm{iso} = \sqrt{3}\sigma^{(0)}$, rather than an improvement in predictive accuracy. To place these results in context, we compare the accuracy of the isotropic chemical shielding predictions against published machine learning models for NMR chemical shifts in related materials. Cuny \textit{et al.}\cite{cuny2016InitioQuality} achieved MAEs of 1.3 and 2.2~ppm for $^{29}$Si and $^{17}$O chemical shifts, respectively, using a neural network model trained on a narrow dataset of SiO$_2$ polymorphs. For aluminosilicate zeolites, which typically exhibit a broader chemical and configurational space, we have recently shown that LASSO-based regression can predict $^{23}$Na and $^{27}$Al isotropic shifts with MAEs of 2.8 and 1.3~ppm, respectively.\cite{lei2023needoperandoa,lei2025machinelearningb} We have also shown that SOAP-based representations combined with KRR outperform LASSO-based models, achieving MAEs typically below 1.5~ppm for $^{27}$Al isotropic shifts across a wide range of zeolite frameworks, Si/Al ratios, and hydration levels.\cite{willimetz2025AluminumSiting,willimetz202527NMRa} On similarly diverse datasets of aluminosilicate glasses, Chaker \textit{et al.}\cite{chaker2019NMRshiftsa} reported MAEs of 1.8~ppm ($^{29}$Si), 3.5~ppm ($^{17}$O), 1.5~ppm ($^{23}$Na), and 1.8~ppm ($^{27}$Al) for isotropic magnetic shielding using SOAP descriptors with linear ridge regression. Our model achieves comparable or better accuracy for the isotropic component across all five nuclei simultaneously, while also providing access to the full tensorial information ($\sigma^{(1)}$ and $\sigma^{(2)}$ components) that scalar models cannot capture. Among existing models that do predict tensorial NMR properties, Venetos \textit{et al.}\cite{venetos2023MachineLearning} achieved an average MAE of 2.82~ppm for $^{29}$Si chemical shifts derived from the symmetric part of the chemical shift tensor, using a model trained on various silicates.

The span $\Omega$, which quantifies the breadth of the chemical shift anisotropy, is predicted with MAEs of 0.69~ppm ($^{1}$H), 3.20~ppm ($^{17}$O), 6.04~ppm ($^{23}$Na), 3.28~ppm ($^{27}$Al), and 1.86~ppm ($^{29}$Si). The skew $\kappa$, a dimensionless parameter bounded between $-1$ and $+1$ that describes the shape of the CSA pattern, is reproduced with MAEs $\leq 0.06$ for all nuclei except $^{23}$Na (MAE = 0.18), consistent with the lower prediction accuracy observed for sodium's spherical tensor components. Overall, these results demonstrate that the model captures both the magnitude and shape of the magnetic shielding anisotropy across all five nuclei.

\begin{figure}[h!]
  \centering
  \includegraphics[width=0.9\textwidth]{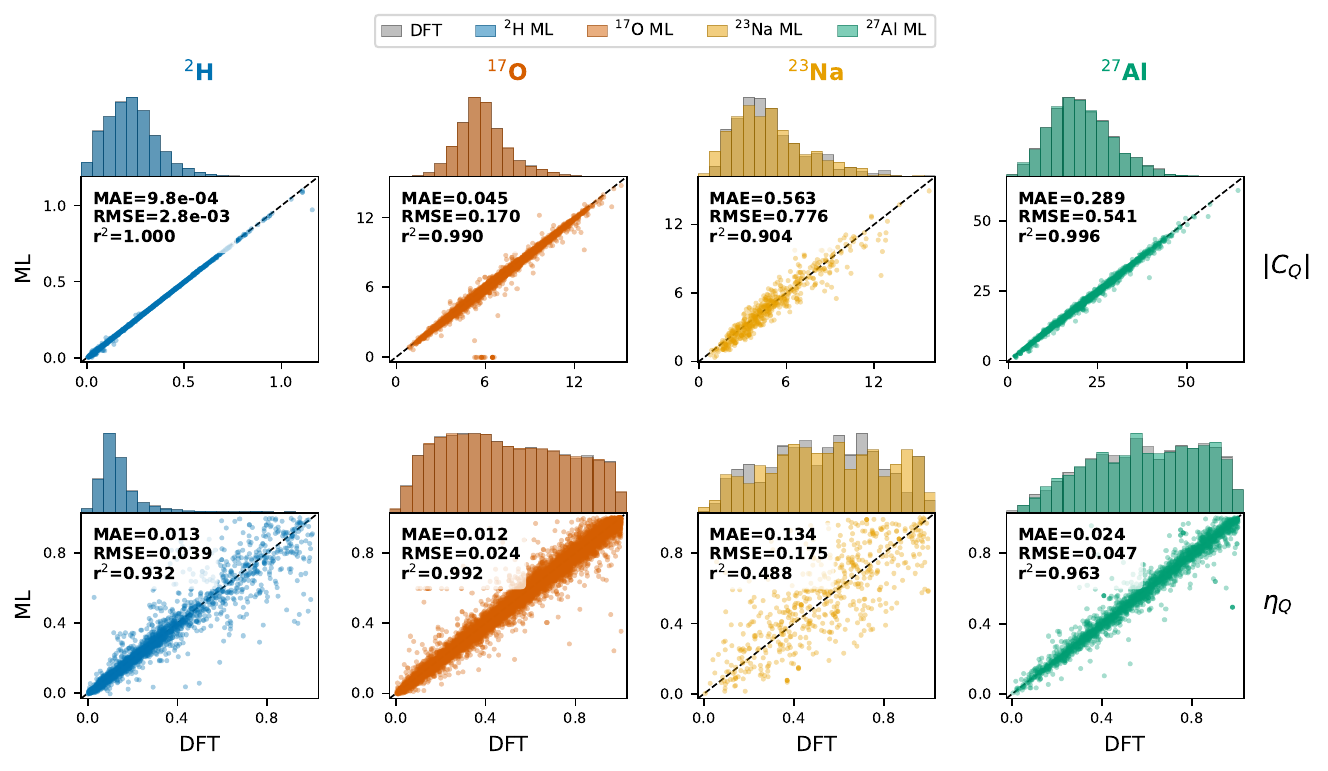}
  \caption{Parity plots comparing DFT-calculated and ML-predicted EFG tensor observables for $^{2}$H, $^{17}$O, $^{23}$Na, and $^{27}$Al on the test set. The absolute value of the quadrupolar coupling constant ($|C_Q|$, in MHz) and the asymmetry parameter ($\eta_Q$, dimensionless) are shown. $^{29}$Si is omitted as it is a spin-$\nicefrac{1}{2}$ nucleus.}
  \label{fig:test-parity-observables}
\end{figure}

\FloatBarrier
For the EFG-derived observables (Figure~\ref{fig:test-parity-observables}), we report $|C_Q|$ and $\eta_Q$ for the four quadrupolar nuclei in our dataset: $^{2}$H, $^{17}$O, $^{23}$Na, and $^{27}$Al. Note that $^{29}$Si (spin-1/2) does not possess a quadrupolar interaction and is therefore absent from this analysis, while $^{2}$H (spin-1) replaces $^{1}$H as the relevant quadrupolar hydrogen isotope. The model predicts $|C_Q|$ with MAEs of 0.00~MHz ($^{2}$H), 0.05~MHz ($^{17}$O), 0.55~MHz ($^{23}$Na), and 0.36~MHz ($^{27}$Al), with $R^2 \geq 0.90$ for all nuclei. The asymmetry parameter $\eta_Q$ is reproduced with MAEs of 0.01 ($^{2}$H), 0.01 ($^{17}$O), 0.14 ($^{23}$Na), and 0.03 ($^{27}$Al). While the prediction of $\eta_Q$ for $^{23}$Na is the least accurate ($R^2 = 0.52$), this again reflects the limited number of sodium training points and the broad distribution of local environments sampled by extra-framework cations.

An important consideration for the EFG predictions is the sign of $C_Q$. Since $C_Q$ is proportional to $V_{zz}$, its sign depends on the eigenvalue ordering. When predicted eigenvalues are close in magnitude, small errors can flip which eigenvalue is assigned as $V_{zz}$, causing a discontinuous sign change in $C_Q$. This discontinuity is a recognized challenge for ML prediction of quadrupolar parameters: Charpentier\cite{charpentier2025FirstprinciplesNMR} showed that scalar regression models fail to predict $C_Q$ and $\eta_Q$ directly due to this effect, recommending instead the prediction of the full EFG tensor, in line with the findings of Harper \textit{et al.}\cite{harper2025Performancemetrics} who demonstrated the superiority of tensorial over scalar learning of EFG-derived observables. Our approach of predicting the full $V^{(2)}$ tensor and deriving $C_Q$ and $\eta_Q$ post-hoc avoids learning discontinuous targets. Inspection of the signed $C_Q$ parity plots (Figure~\ref{fig_si:test-parity-efg-observables}) reveals that sign errors persist in the derived observables, particularly for $^{23}$Na (MAE = 1.52~MHz, $R^2 = 0.63$) and $^{27}$Al (MAE = 1.00~MHz, $R^2 = 0.94$), while the signed prediction for $^{17}$O remains accurate (MAE = 0.12~MHz, $R^2 = 0.97$). From the experimental perspective, however, the sign of $C_Q$ cannot be determined from standard solid-state NMR experiments, as spectra are invariant to the sign of the quadrupolar coupling.\cite{man2011QuadrupolarInteractions} Consequently, the use of $|C_Q|$ in Figure~\ref{fig:test-parity-observables} reflects experimentally meaningful accuracy, and the sign ambiguity is not expected affect the practical utility of the model for spectral prediction and interpretation.

\FloatBarrier
\subsection{Transferability to unseen zeolite frameworks}
A key advantage of predicting the full NMR tensor rather than individual scalar observables, is the ability to directly simulate realistic solid-state NMR spectra. For quadrupolar nuclei, accurate spectral simulation requires not only the magnitude of the interactions but also the relative orientation between the magnetic shielding and EFG tensors, information that is only accessible from the full tensors. To demonstrate this capability and assess the transferability of our model, we simulated $^{29}$Si and $^{27}$Al MAS NMR spectra for the RTH zeolite, a framework not included in the training dataset, and compared them against both DFT-simulated spectra and experimental data recently published.\cite{willimetz202527NMRa,willimetz2025AluminumSiting}

\begin{figure}[h!]
    \centering
    \includegraphics[width=0.8\textwidth]{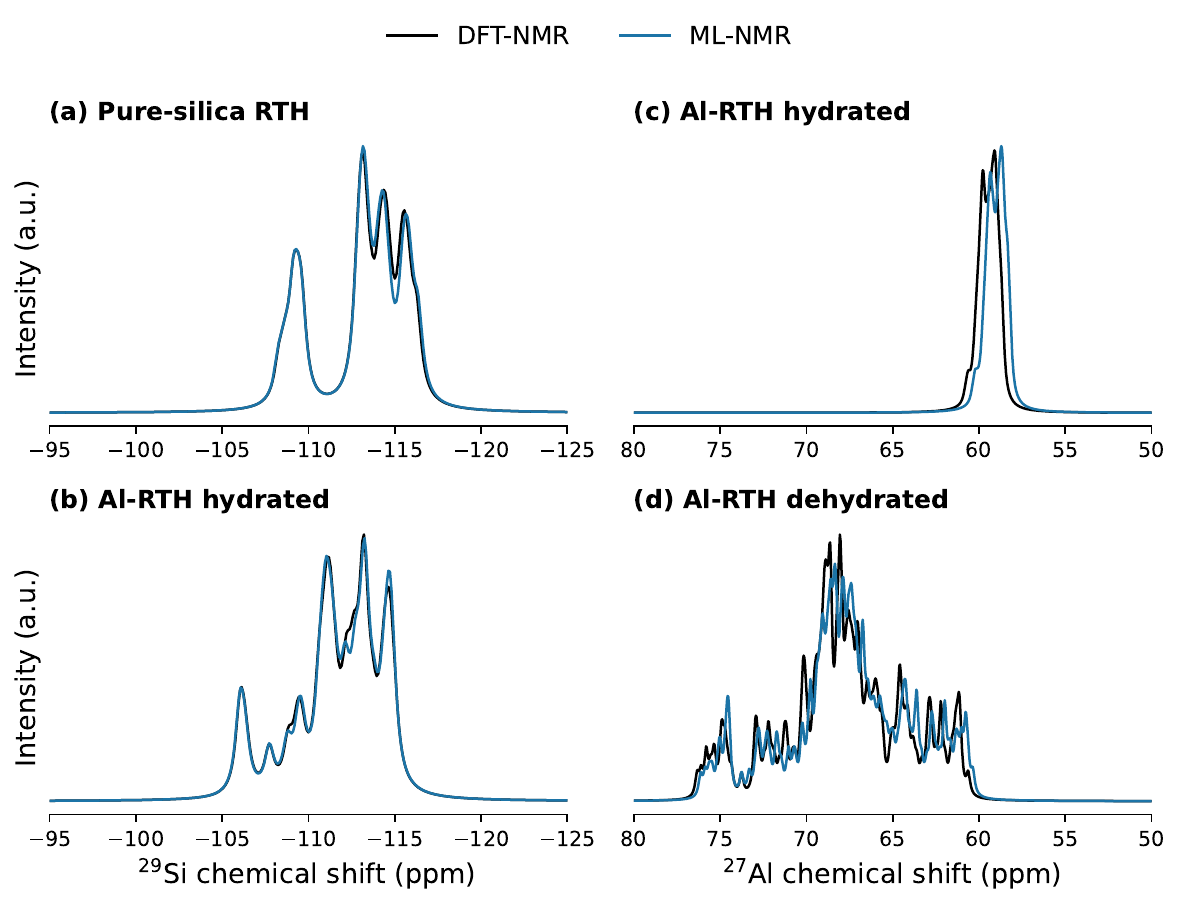}
    \caption{Simulated $^{29}$Si NMR spectrum of a) pure-silica RTH and b) Al-containing RTH zeolites, $^{27}$Al NMR spectra of Al-containing RTH zeolite in c) hydrated and d) dehydrated form.}
    \label{fig:rth-nmr}
\end{figure}

Figure~\ref{fig:rth-nmr} shows the simulated spectra for siliceous RTH, and for hydrated and dehydrated Al-containing H-RTH. For the siliceous form, the ML-predicted $^{29}$Si spectrum correctly resolves all crystallographically distinct T sites, with isotropic chemical shifts in close agreement with both DFT and experiment (Table~\ref{tab:rth-si}). Upon aluminum incorporation, the $^{29}$Si shows peaks from Q$^4$(0Al) and Q$^4$(1Al) sites, with the latter appearing at higher chemical shifts, and we observe resonance over a wider range due to the distribution of local environments arising from insertion of Al in different T-sites. For $^{27}$Al, the ML-predicted spectra capture both the sharp resonance observed under hydrated conditions and the characteristic quadrupolar brodening upon dehydration, resulting from distortion to the aluminum coordination. For siliceous RTH, the ML-predicted $^{29}$Si isotropic chemical shifts agree with DFT to within 0.1~ppm and with experiment to within 0.5~ppm for all four T sites (Table~\ref{tab:rth-si}). For hydrated Al-containing H-RTH, the $^{27}$Al isotropic shifts are reproduced with similar accuracy (Table~\ref{tab:rth-al-hyd}), while the predicted $C_Q$ values faithfully reproduce the DFT reference but overestimate the experimental values, likely reflecting limitations of the reference level and short timescale of the ML-driven simulation. For the dehydrated form, both DFT and ML predict substantially larger $C_Q$ values (17--18 MHz, Table~\ref{tab:rth-al-dehyd}), consistent with the characteristic broadening of the $^{27}$Al spectrum observed upon dehydration. These results demonstrate that the model generalizes reliably to zeolite frameworks outside its training distribution and can be used to simulate NMR spectra under realistic conditions, beyond the small-scale, necessarily simplified models that are accessible to direct first-principles predictions.

\begin{table}[ht!]
    \centering
    \caption{Comparison between DFT-calculated, ML-predicted, and experimental $^{29}$Si isotropic chemical shifts ($\delta_\mathrm{iso}$) for the crystallographically distinct T sites in siliceous RTH. DFT and ML values were obtained from time-averaged from 1~ns MD trajectories. Experimental values from Willimetz et al.\cite{willimetz2025AluminumSiting}}
    \begin{tabular}{c c c c}
        \toprule
        T-site & DFT (ppm)  & ML (ppm)  & Experimental (ppm) \\
        \midrule
        T1     & -115.74 & -115.84 & -115.5 \\
        T2     & -113.44 & -113.47 & -113.6 \\
        T3     & -113.93 & -113.92 & -113.4 \\
        T4     & -109.14 & -109.13 & -108.7 \\
        \bottomrule
    \end{tabular}
    \label{tab:rth-si}
\end{table}

\begin{table}[ht!]
    \centering
    \caption{Comparison of DFT-calculated, ML-predicted, and experimental $^{27}$Al isotropic chemical shifts ($\delta_\mathrm{iso}$) and quadrupolar coupling constants ($C_Q$) for hydrated Al-containing H-RTH. Values are time-averaged from 1~ns ML-MD trajectories. Experimental values from Willimetz et al.\cite{willimetz2025AluminumSiting}.}
    \begin{tabular}{c c c c c c c}
        \toprule
        \multirow{2}{*}{T-site}  & \multicolumn{2}{c}{DFT}     & \multicolumn{2}{c}{ML}      & \multicolumn{2}{c}{Experimental} \\
        \cmidrule(l{3pt}r{3pt}){2-3}
        \cmidrule(l{3pt}r{3pt}l{3pt}r{3pt}){4-5}
        \cmidrule(l{3pt}r{3pt}){6-7}
                & $\delta_{iso}$ (ppm)    & $C_Q$ (MHz) & $\delta_{iso}$ (ppm)    & $C_Q$ (MHz) & $\delta_{iso}$ (ppm)    & $C_Q$ (MHz) \\
        \midrule
        T1  & 57.96 & 3.81  & 57.41 & 3.74  & 58.1  & 1.4   \\
        T2  & 58.25	& 2.75	& 57.87	& 2.73	& 57.2	& 1.8  \\  
        T3	& 55.22	& 3.91	& 55.02	& 3.91	& 56.6	& 1.9   \\   
        T4	& 62.45	& 0.55	& 61.84	& 0.66	& 61.2	& 1.7   \\   
        \bottomrule
    \end{tabular}
    \label{tab:rth-al-hyd}
\end{table}

\FloatBarrier
\section{Conclusions and outlook}

In this work we developed a MACE-based equivariant graph neural network model for predicting NMR magnetic shielding and electric field gradient tensors across the broad chemical and configurational space of aluminosilicate zeolites. Our models simultaneously predict quantities for six NMR-active nuclei ($^{1}$H, $^{2}$H, $^{17}$O, $^{23}$Na, $^{27}$Al, and $^{29}$Si) spanning diverse Si/Al ratios, water contents, and H$^+$/Na$^+$ charge-balancing cations, going significantly beyond prior ML-NMR approaches that are typically restricted to one or two nuclei in datasets with narrower configuration spaces. The transferability of the model was validated for RTH zeolite, a framework not included in the training dataset, where ML-predicted $^{29}$Si and $^{27}$Al NMR spectra closely reproduce both DFT-calculated references and experimental measurements.\cite{willimetz2025AluminumSiting} This demonstrates the model's ability to generalize to unseen zeolite frameworks and to bridge the gap between atomistic simulations and real-world NMR experiments.

Unlike most existing ML-NMR models, which typically predict only isotropic chemical shifts, our model predicts the full irreducible spherical tensor components, $\sigma^{(0)}$, $\sigma^{(1)}$, and $\sigma^{(2)}$ for magnetic shielding and $V^{(2)}$ for EFG, providing access to the complete set of NMR observables such isotropic shifts, chemical shift anisotropy parameters ($\Omega$, $\kappa$), and quadrupolar parameters ($C_Q$, $\eta_Q$). We showed that EFG tensors are easier to learn, and also easier to compute from DFT, than magnetic shielding tensors, making EFG predictions an accessible target for other material classes as well as other spectroscopies such as M{\"o}ssbauer and nuclear quadrupole resonance. 


While MLIPs are now a mainstream technique for materials simulations, comparable ML models for tensorial NMR parameters are at a much earlier stage of development. 
In the same way that MLIPs have evolved from specialised tools to widespread use, can tensor NMR models become commonplace? With the example of an important class of inorganic materials, our study highlights some more general directions for future research towards that goal. (i)~We showed that it is beneficial, or even necessary, to remove outliers from the training data (Tables \ref{tab:iqr-l0-ms}--\ref{tab:iqr-l2-efg}); can this be developed into more general protocols? Contrary to MLIP training, where energy and force labels remain physically meaningful for any geometry, NMR tensors of highly distorted structures can take extreme values that do not correspond to experimental observables and degrade model accuracy. (ii)~What is the reason for the remaining differences, say, in $C_{Q}$, between the reference DFT level and experiment (Table~\ref{tab:rth-al-hyd}), and will higher-level DFT be able to help with addressing this discrepancy?\cite{windeck2023spectroscopic} (iii)~What would it take to develop a `foundational' NMR model for inorganic materials? In the case of MLIPs, wide swathes of the Periodic Table are covered in (pre-) training datasets such as MPtrj, \cite{deng2023chgnet} Alexandria, \cite{schmidt2024improving} or OMat24; \cite{barroso2024open} in contrast, for NMR models, many elements and compounds are metallic and therefore not easily accessible. Would a foundational NMR model perhaps include all NMR-active nuclei or focus only on those that are commonly measured in the laboratory? The higher computational cost of DFT-NMR relative to single-point calculation further constrains the generation large datasets. And would it start from the same type of structures as MLIP models do, or would it require bespoke structure generation? (iv)~Can automation help, in the same way that MLIP training dataset generation can be (at least partly) automated? \cite{zhang2020dp,  menon2024electrons, liu2025automated} (v)~Can pre-training and fine-tuning strategies, which have proven highly beneficial for MLIPs, \cite{smith2019approaching, gardner2024synthetic, kaur2025data} be implemented and used for NMR models as well? (vi)~Can ML-NMR models be coupled with MLIPs for on-the-fly prediction of NMR parameters during molecular dynamics simulations?

\section*{Acknowledgement}
CB, LG and CJH acknowledge the support of Czech Science Foundation (26-23277S) and Charles University Centre of Advanced Materials (CUCAM, OP VVV Excellent Research Teams, project number CZ.02.1.01/0.0/0.0/15\_003/0000417). The computations at Charles University were supported by the Ministry of Education, Youth and Sports of the Czech Republic through the e-INFRA CZ (ID:90254). CB acknowledges the funding from the European Union's Horizon Europe research and innovation program under the ERA-PF grant agreement no. 101180584.
This work was supported by UK Research and Innovation [grant number EP/X016188/1].

\section*{Data availability}
Data, models and code to reproduce the results will be made available at \href{https://github.com/Nanomaterials-Modelling-Group-Prague/ZeoMLNMR}{github.com/Nanomaterials-Modelling-Group-Prague/ZeoMLNMR}.

\nocite{mason1993conventions,anet1990NMRrelaxation,autschbach2010analysis,benmahmoud2025Graphneuralnetworkpredictions}

\bibliography{references}

\clearpage
\setcounter{figure}{0}
\setcounter{table}{0}
\setcounter{equation}{0}
\setcounter{section}{0}
\renewcommand{\thefigure}{S\arabic{figure}}
\renewcommand{\thetable}{S\arabic{table}}
\renewcommand{\theequation}{S\arabic{equation}}
\renewcommand{\thesection}{S\arabic{section}}
\renewcommand{\theHfigure}{SI.\arabic{figure}}
\renewcommand{\theHtable}{SI.\arabic{table}}
\renewcommand{\theHequation}{SI.\arabic{equation}}
\renewcommand{\theHsection}{SI.\arabic{section}}

\input{si}

\end{document}

%% file: si.tex

\section*{Supporting Information}

\vspace{-1.55em}
{\let\clearpage\relax \listoffigures}
\listoftables

\clearpage

\section{NMR tensor and irreducible spherical tensor decomposition}

The magnetic shielding tensor $\sigma$ is a rank-2 Cartesian tensor that relates the induced magnetic field at a nucleus to the static magnetic field, and is represented as a $3\times3$ matrix with 9 independent components:
\begin{equation}
    \sigma = \begin{bmatrix}
    \sigma_{xx} & \sigma_{xy} & \sigma_{xz} \\
    \sigma_{yx} & \sigma_{yy} & \sigma_{yz} \\
    \sigma_{zx} & \sigma_{zy} & \sigma_{zz}
\end{bmatrix}.
\end{equation}
In general, $\sigma$ is non-symmetric tensor. Its isotropic part, $\sigma_\mathrm{iso} = \frac{1}{3}\mathrm{Tr}(\sigma)$, dictates the position of the NMR resonance. The symmetric part of $\sigma$, in the principal axis system (PAS) where it is diagonal with eigenvalues ordered $\sigma_{11} \ge \sigma_{22} \ge \sigma_{33}$, is fully characterized by $\sigma_\mathrm{iso}$, the span $\Omega = \sigma_{33} - \sigma_{11} \geq 0$, and the skew $\kappa = 3(\sigma_{22} - \sigma_\mathrm{iso})/\Omega$, which together parameterize the chemical shift anisotropy (CSA).\cite{mason1993conventions} The antisymmetric part often is discared, as it does not contribute to the NMR spectrum.\cite{anet1990NMRrelaxation}

The electric field gradient (EFG) tensor $V$ describes the second spatial derivative of the electrostatic potential of a nucleus, and can be represented as:

\begin{equation}
V = \begin{bmatrix}
    V_{xx} & V_{xy} & V_{xz} \\
    V_{xy} & V_{yy} & V_{yz} \\
    V_{xz} & V_{yz} & V_{zz}
\end{bmatrix}, \qquad V_{xx} + V_{yy} + V_{zz} = 0.
\end{equation}

Unlike a magnetic shielding tensor $\sigma$, the EFG tensor $V$ is by definition both symmetric and traceless, leaving only 5 independent components. In the PAS, with eigenvalues ordered $|V_{zz}| \geq |V_{yy}| \geq |V_{xx}|$, the EFG is typically characterized by the quadrupolar coupling constant $C_Q = eQV_{zz}/h$ and the asymmetry parameter $\eta_Q = (V_{xx} - V_{yy})/V_{zz}$, where $e$ is the elementary charge, $Q$ the nuclear quadrupole moment, and $h$ Planck's constant.\cite{autschbach2010analysis}

Rank-2 Cartesian tensors can be transformed into speherical tensors and then decomposed into irreducible representations of the rotation group SO(3) via the Clebsch--Gordan series $ \mathbf{0} \oplus \mathbf{1} \oplus \mathbf{2}$, yielding components that transform as real spherical harmonics of degree $\ell = 0$, $1$, and $2$. Consequently, the magnetic shielding tensor $\sigma$ can be decomposed as:
\begin{equation}
    \sigma = \sigma^{(0)} \oplus \sigma^{(1)} \oplus \sigma^{(2)},
\end{equation}
where $\sigma^{(0)}$ (1 term) and $\sigma^{(2)}$ (5 terms) encode the symmetric isotropic and symmetric traceless parts of the tensor, respectively, while $\sigma^{(1)}$ (3 terms) encodes the antisymmetric part. Since, the EFG tensor $V$ is already symmetric and traceless, it can be decomposed as:
\begin{equation}
    V = V^{(2)}.
\end{equation}

Both $\sigma$ and $V$ are even-parity tensors ($p = +1$ under spatial inversion), so all their irreducible components carry even parity.\cite{benmahmoud2025Graphneuralnetworkpredictions}

\section{Convergence of DFT NMR parameters}

The sensitivity of key NMR parameters, namely $\sigma_{iso}$, $\Omega$, ($\kappa$), $C_Q$ and $\eta_Q$,  parameters to the plane-wave basis set and Brillouin-zone sampling, systematic convergence tests were performed across 10 diverse structures. For each structure, GIPAW DFT calculations were carried out at plane-wave cutoff energies between 400 and 1200 eV, with k-point spacing fixed at 0.05 \AA$^{-1}$, and at k-point spacings between 0.09 and 0.04 \AA$^{-1}$, with cutoff fixed at 600 eV. Convergence was quantified as the root-mean-square deviation (RMSD) of each parameter relative to the most converged calculation, cutoff = 1200 eV and k-spacing = 0.04 \AA$^{-1}$. The RMSD was computed for all atoms of a given element within each structure.

\begin{figure}[ht!]
    \centering
    \includegraphics[width=\textwidth]{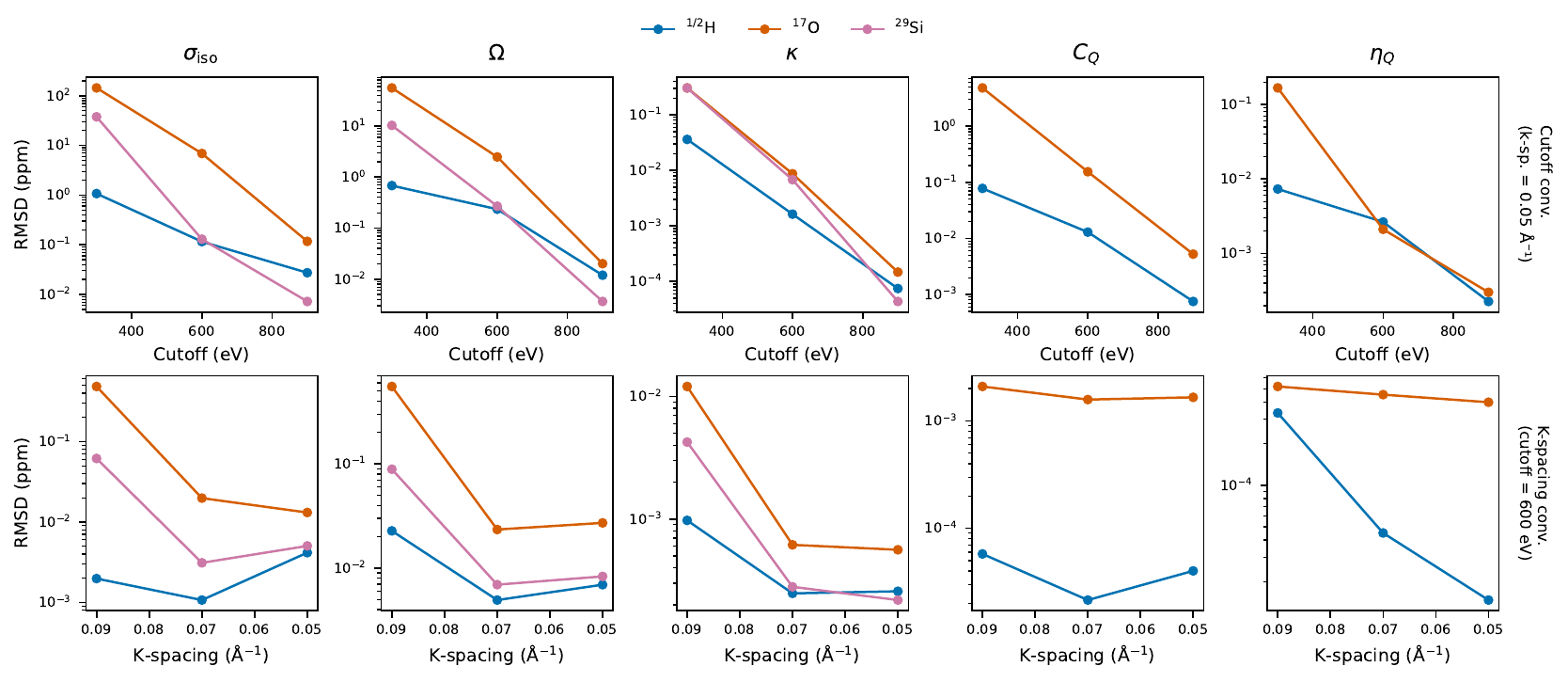}
    \caption{Convergence of DFT-computed NMR parameters of a dehydrated Na-exchanged aluminosilicate zeolites (structure 1).}
    \label{fig:dft-conv1}
\end{figure}

\begin{figure}[ht!]
    \centering
    \includegraphics[width=\textwidth]{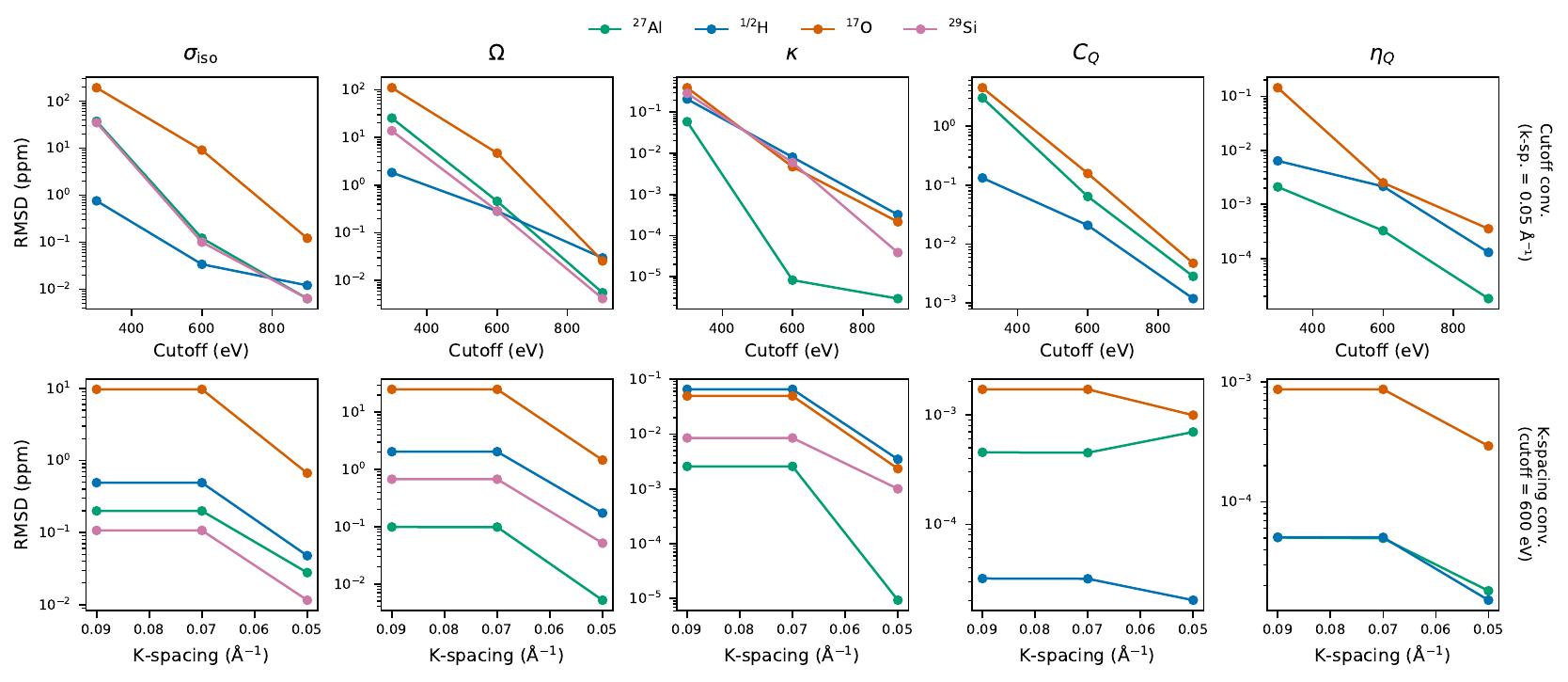}
    \caption{Convergence of DFT-computed NMR parameters of a dehydrated H-exchanged aluminosilicate zeolites (structure 2).}
    \label{fig:dft-conv2}
\end{figure}

\begin{figure}[ht!]
    \centering
    \includegraphics[width=\textwidth]{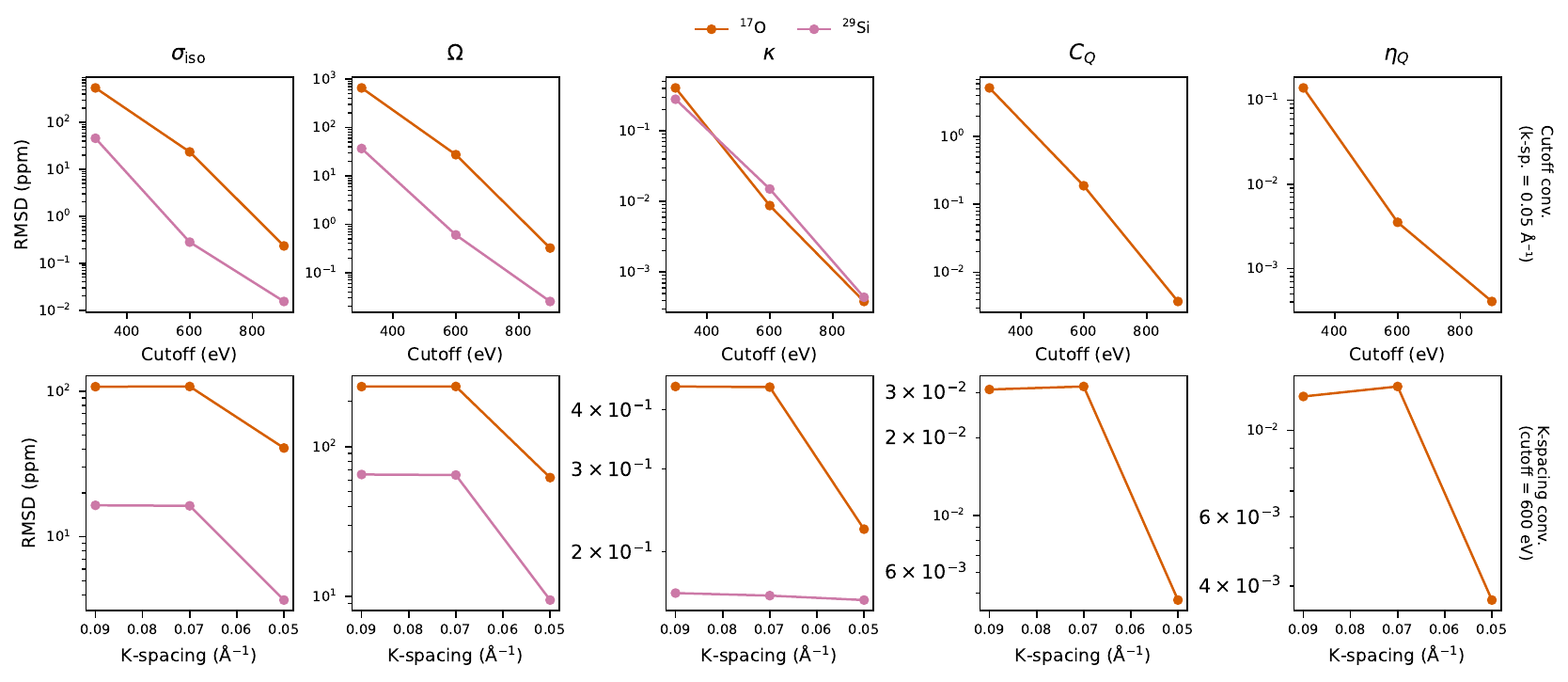}
    \caption{Convergence of DFT-computed NMR parameters of a bulk Na-aluminosilicate (structure 3).}
    \label{fig:dft-conv3}
\end{figure}

\begin{figure}[ht!]
    \centering
    \includegraphics[width=\textwidth]{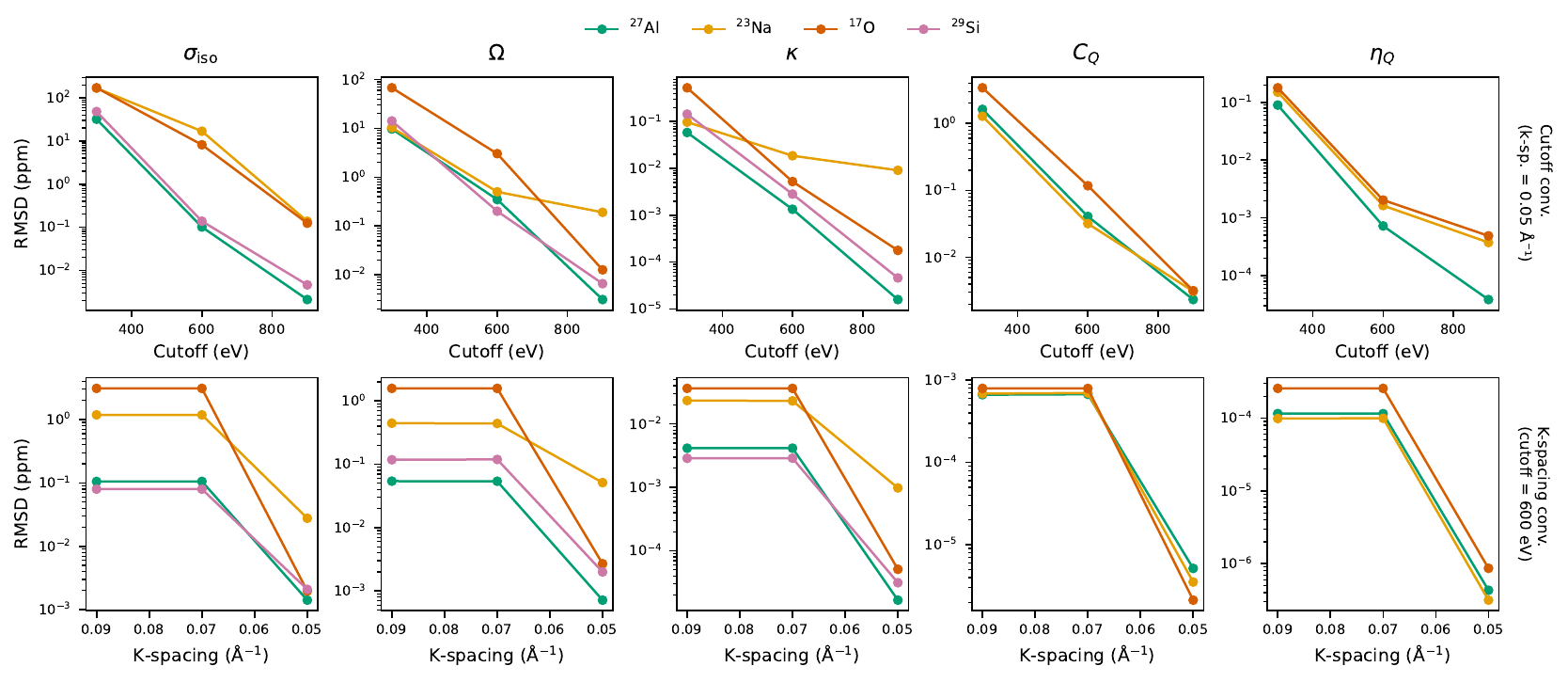}
    \caption{Convergence of DFT-computed NMR parameters of a Al$_2$O$_3$ polymorph (structure 4).}
    \label{fig:dft-conv4}
\end{figure}

\begin{figure}[ht!]
    \centering
    \includegraphics[width=\textwidth]{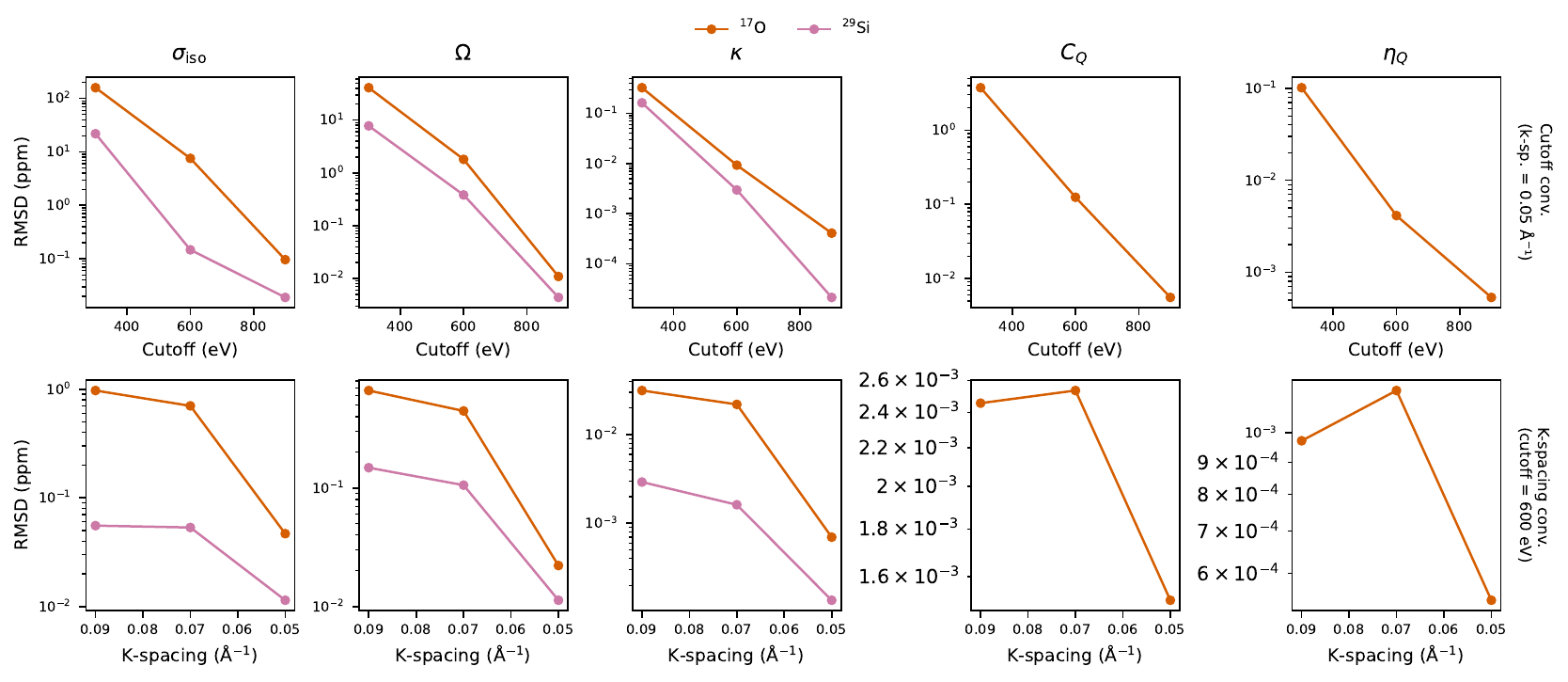}
    \caption{Convergence of DFT-computed NMR parameters of a SiO$_2$ polymorph (structure 5).}
    \label{fig:dft-conv5}
\end{figure}

\begin{figure}[ht!]
    \centering
    \includegraphics[width=\textwidth]{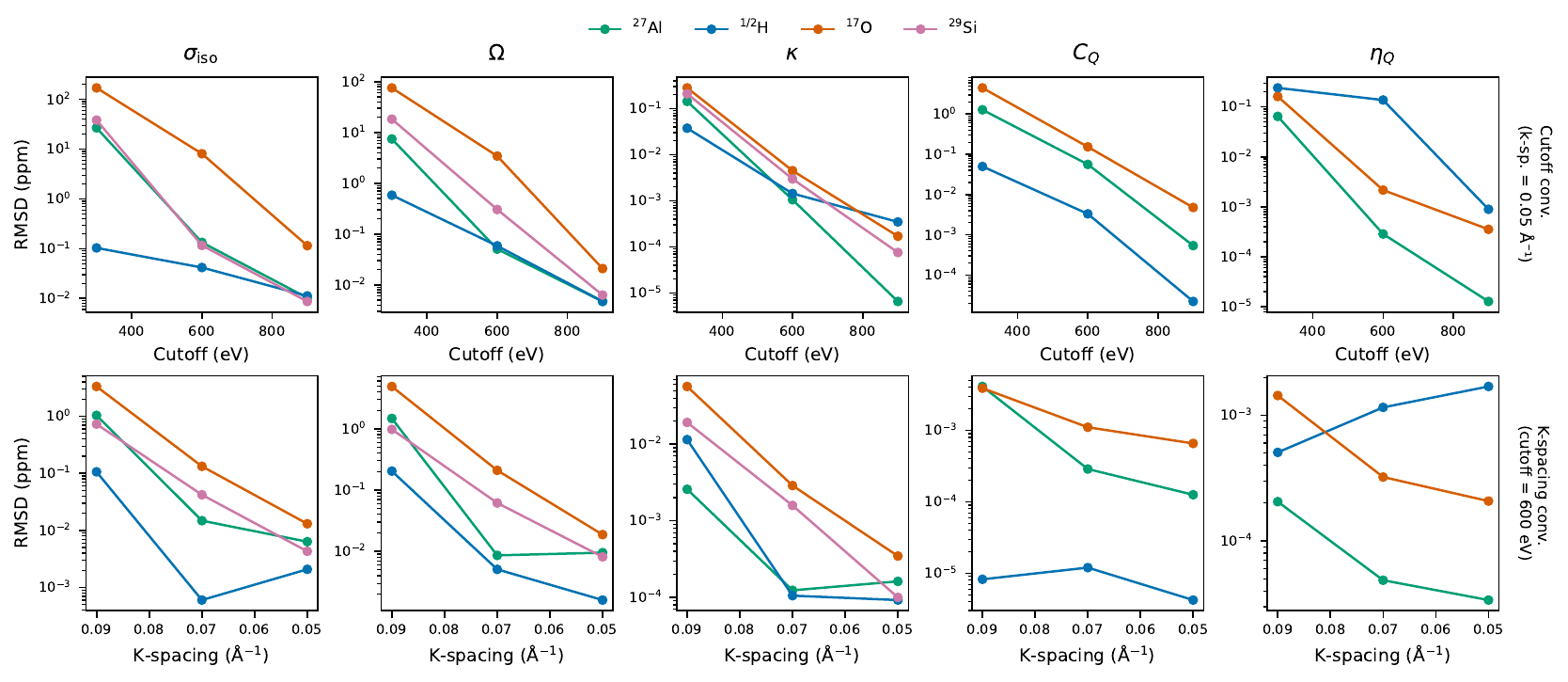}
    \caption{Convergence of DFT-computed NMR parameters of a defective Na-exchanged aluminosilicate zeolites (structure 6).}
    \label{fig:dft-conv6}
\end{figure}

\begin{figure}[ht!]
    \centering
    \includegraphics[width=\textwidth]{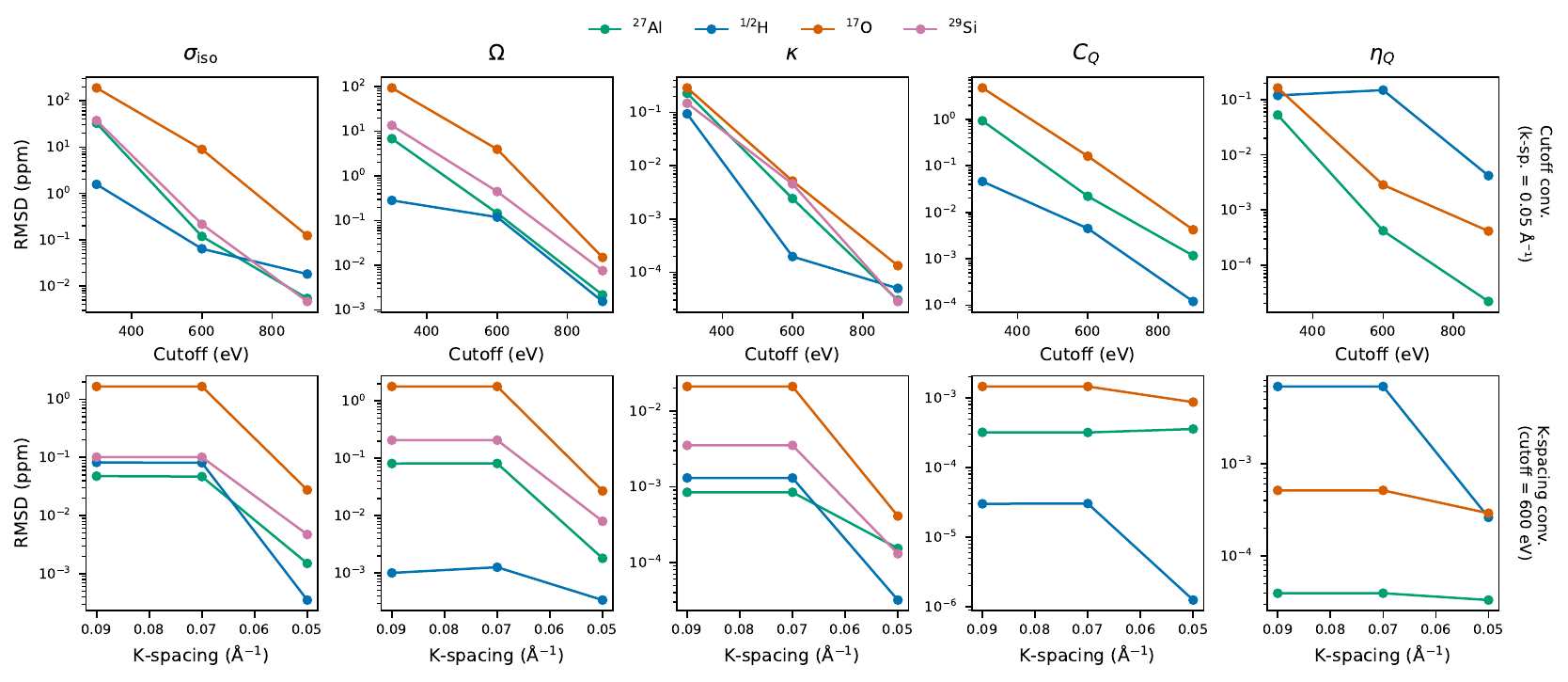}
    \caption{Convergence of DFT-computed NMR parameters of a siliceous zeolite (structure 7).}
    \label{fig:dft-conv7}
\end{figure}

\begin{figure}[ht!]
    \centering
    \includegraphics[width=\textwidth]{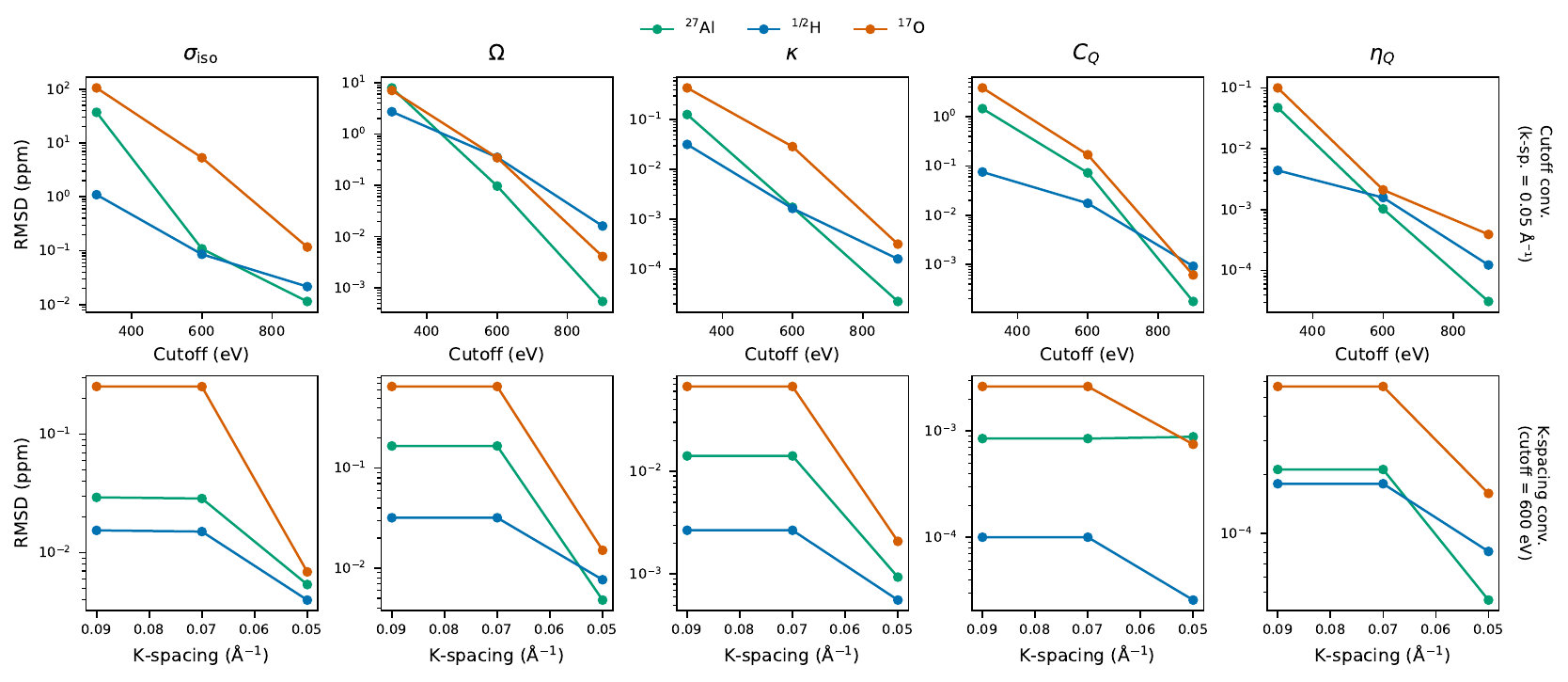}
    \caption{Convergence of DFT-computed NMR parameters of water clusters (structure 8).}
    \label{fig:dft-conv8}
\end{figure}

\begin{figure}[ht!]
    \centering
    \includegraphics[width=\textwidth]{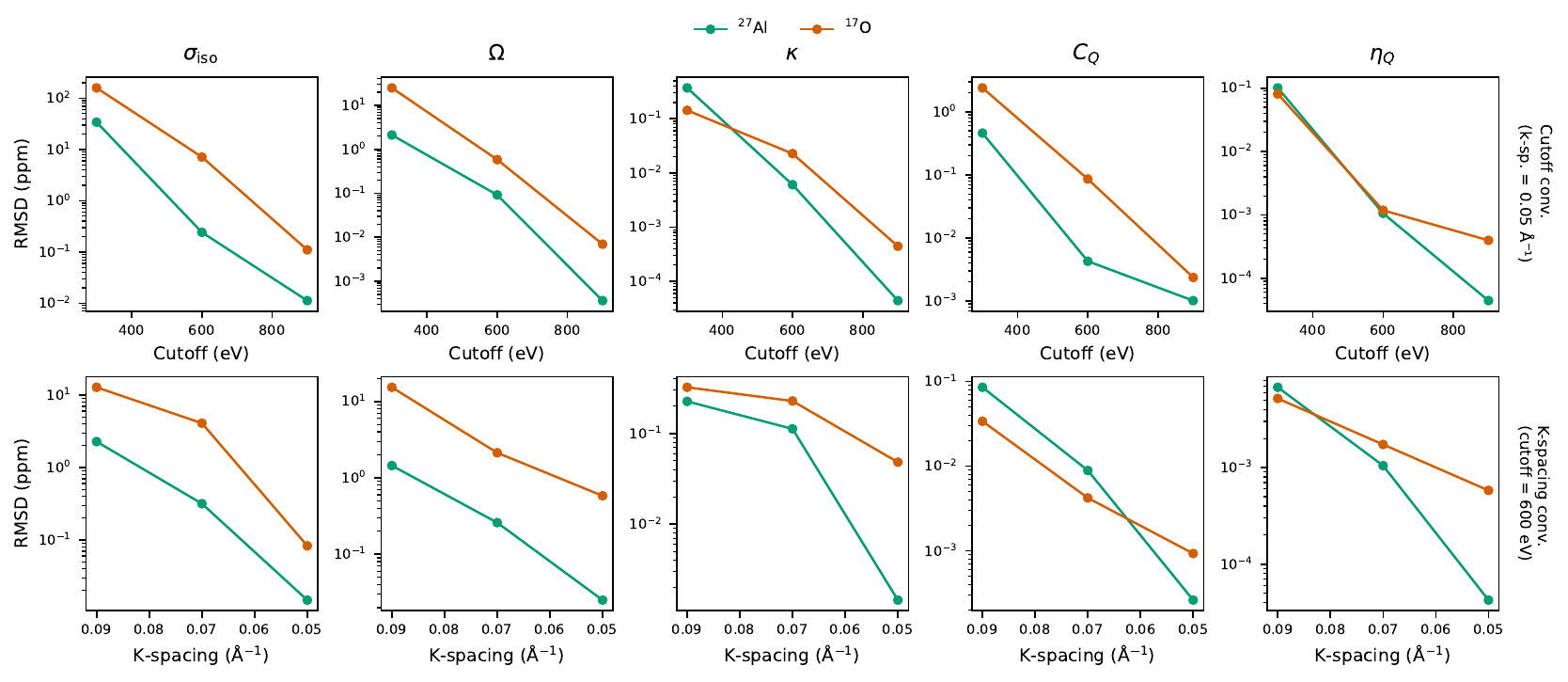}
    \caption{Convergence of DFT-computed NMR parameters of a dehydrated high Al-containing H-zeolite (structure 9).}
    \label{fig:dft-conv9}
\end{figure}

\begin{figure}[ht!]
    \centering
    \includegraphics[width=\textwidth]{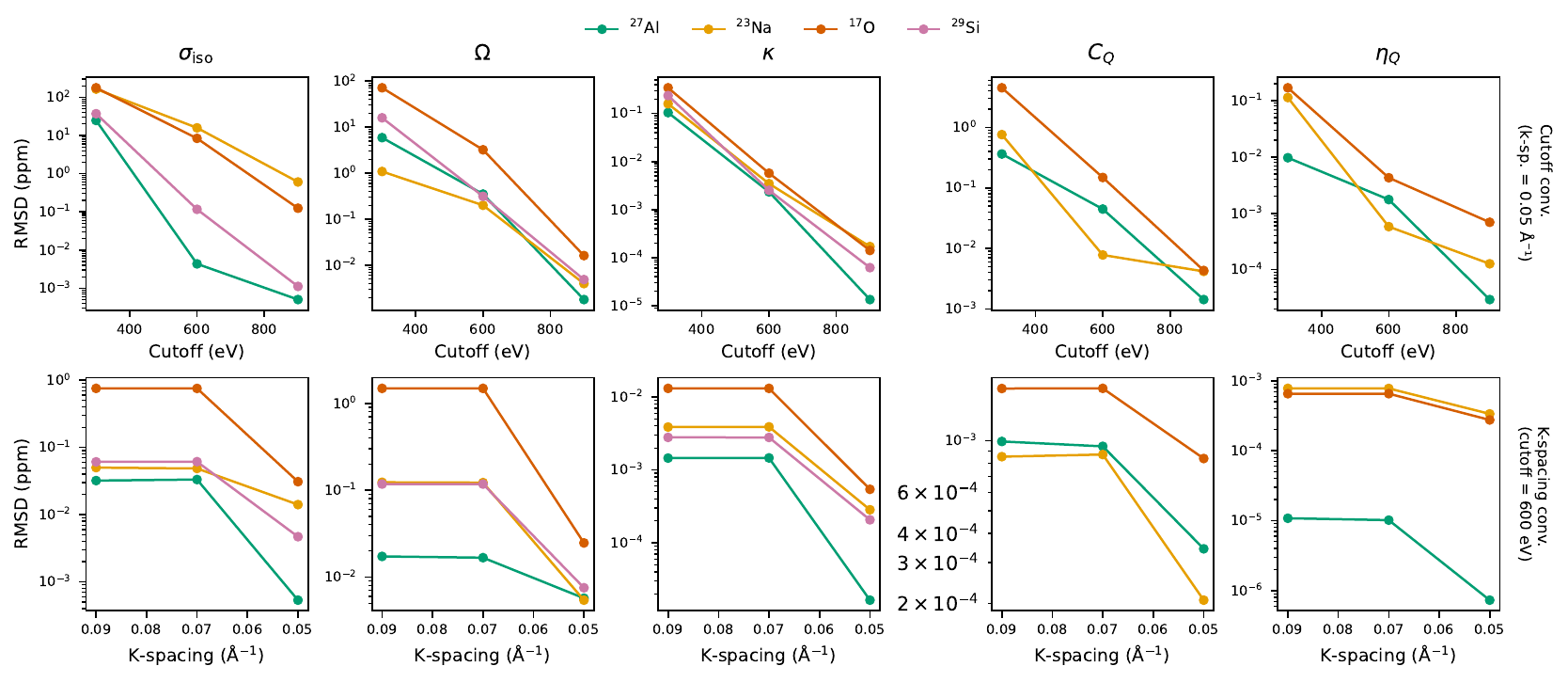}
    \caption{Convergence of DFT-computed NMR parameters of a aluminium hydroxide polymorph (structure 10).}
    \label{fig:dft-conv10}
\end{figure}

\FloatBarrier
\section{Selection of dataset}

\begin{table}[ht!]
    \centering
    \caption{Comparison of MAE and RMSE for $\sigma^{(0)}$ in the test sets of the initial dataset and those cleaned with IQR with scale factors k = 2, 3, and 4.}
    \resizebox{\textwidth}{!}{
    \begin{tabular}{c c c c c c c c c c c c c c c c}
    \toprule
        \multirow{2}{*}{Dataset} & \multicolumn{3}{c}{$^{1}$H} & \multicolumn{3}{c}{$^{17}$O} & \multicolumn{3}{c}{$^{23}$Na} & \multicolumn{3}{c}{$^{27}$Al} & \multicolumn{3}{c}{$^{29}$Si} \\
        \cmidrule(l{3pt}r{3pt}){2-4}
        \cmidrule(l{3pt}r{3pt}){5-7}
        \cmidrule(l{3pt}r{3pt}){8-10}
        \cmidrule(l{3pt}r{3pt}){11-13}
        \cmidrule(l{3pt}r{3pt}){14-16}
                & {MAE}   & {RMSE} & {\%RMSE}  & {MAE}     & {RMSE}  & {\%RMSE}  & {MAE}   & {RMSE}  & {\%RMSE}  & {MAE}   & {RMSE}  & {\%RMSE} & {MAE}   & {RMSE} & {\%RMSE} \\
        \midrule
        Full    & 1.49 & 5.23 & 40.47 & 11.82   & 164.78    & 56.54 & 7.69  & 13.39 & 29.52 & 4.20  & 8.16  & 25.79 & 3.11  & 7.66  & 20.13 \\
        IQR 4   & 0.60 & 1.05 & 12.70 & 4.47    & 12.43     & 11.52 & 5.37  & 7.68  & 16.31 & 2.14  & 3.60  & 12.02 & 1.57  & 4.08  & 12.23 \\
        IQR 3   & 0.52 & 0.80 & 9.76 & 3.94     & 10.58     & 10.78 & 5.29  & 7.03  & 14.65 & 2.03  & 3.77  & 12.75 & 1.32  & 2.38  & 7.88  \\
        IQR 2   & 0.42 & 0.65 & 8.07 & 2.68     & 5.69      & 7.01  & 5.19  & 7.60  & 18.07 & 1.62  & 2.59  & 9.25  & 0.95  & 1.64  & 6.49  \\
    \bottomrule
    \end{tabular}
    }
    \label{tab:iqr-l0-ms}
\end{table}

\begin{table}[ht!]
    \centering
    \caption{Comparison of MAE and RMSE for $\sigma^{(1)}$ in the test sets of the initial dataset and those cleaned with IQR with scale factors k = 2, 3, and 4.}
    \resizebox{\textwidth}{!}{
    \begin{tabular}{c c c c c c c c c c c c c c c c}

    \toprule
        \multirow{2}{*}{Dataset} & \multicolumn{3}{c}{$^{1}$H} & \multicolumn{3}{c}{$^{17}$O} & \multicolumn{3}{c}{$^{23}$Na} & \multicolumn{3}{c}{$^{27}$Al} & \multicolumn{3}{c}{$^{29}$Si} \\
        \cmidrule(l{3pt}r{3pt}){2-4}
        \cmidrule(l{3pt}r{3pt}){5-7}
        \cmidrule(l{3pt}r{3pt}){8-10}
        \cmidrule(l{3pt}r{3pt}){11-13}
        \cmidrule(l{3pt}r{3pt}){14-16}
                & {MAE}   & {RMSE} & {\%RMSE}  & {MAE}     & {RMSE}  & {\%RMSE}  & {MAE}   & {RMSE}  & {\%RMSE}  & {MAE}   & {RMSE}  & {\%RMSE} & {MAE}   & {RMSE} & {\%RMSE} \\
        \midrule
        Full    & 0.51 & 5.34 & 97.08 & 5.44 & 84.45 & 89.69   & 1.17 & 2.45 & 99.68    & 2.24 & 3.41 & 50.36    & 2.09 & 4.04 & 36.35  \\
        IQR 4   & 0.26 & 0.46 & 39.52 & 2.24 & 7.13  & 73.83   & 0.91 & 1.33 & 86.28    & 1.40 & 2.15 & 33.53    & 1.19 & 7.01 & 66.19  \\
        IQR 3   & 0.24 & 0.37 & 33.08 & 1.92 & 8.23  & 79.86   & 0.78 & 1.09 & 85.23    & 1.33 & 2.13 & 34.29    & 1.04 & 1.75 & 17.30  \\
        IQR 2   & 0.20 & 0.28 & 26.36 & 1.21 & 2.13  & 36.23   & 0.74 & 1.04 & 86.11    & 0.99 & 1.64 & 28.56    & 0.72 & 1.21 & 12.59    \\
        \bottomrule
    \end{tabular}
    }
    \label{tab:iqr-l1-ms}
\end{table}

\begin{table}[ht!]
    \centering
    \caption{Comparison of MAE and RMSE for $\sigma^{(2)}$ in the test sets of the initial dataset and those cleaned with IQR with scale factors k = 2, 3, and 4.}
    \resizebox{\textwidth}{!}{
    \begin{tabular}{c c c c c c c c c c c c c c c c}
    \toprule
        \multirow{2}{*}{Dataset} & \multicolumn{3}{c}{$^{1}$H} & \multicolumn{3}{c}{$^{17}$O} & \multicolumn{3}{c}{$^{23}$Na} & \multicolumn{3}{c}{$^{27}$Al} & \multicolumn{3}{c}{$^{29}$Si} \\
        \cmidrule(l{3pt}r{3pt}){2-4}
        \cmidrule(l{3pt}r{3pt}){5-7}
        \cmidrule(l{3pt}r{3pt}){8-10}
        \cmidrule(l{3pt}r{3pt}){11-13}
        \cmidrule(l{3pt}r{3pt}){14-16}
                & {MAE}   & {RMSE} & {\%RMSE}  & {MAE}     & {RMSE}  & {\%RMSE}  & {MAE}   & {RMSE}  & {\%RMSE}  & {MAE}   & {RMSE}  & {\%RMSE} & {MAE}   & {RMSE} & {\%RMSE} \\
        \midrule
        Full    & 1.28 & 5.14 & 47.20   & 8.42 & 120.67 & 68.82  & 5.00 & 8.06 & 44.25 & 4.54 & 7.80 & 18.58    & 3.42 & 6.58 & 16.34    \\
        IQR 4   & 0.55 & 1.30 & 16.67   & 3.36 & 10.12  & 27.48  & 3.48 & 5.01 & 28.08 & 2.52 & 4.55 & 11.18    & 1.81 & 7.76 & 20.13    \\
        IQR 3   & 0.49 & 0.82 & 10.41   & 2.93 & 8.69   & 25.09  & 3.27 & 4.52 & 27.12 & 2.45 & 4.06 & 10.48    & 1.62 & 2.79 & 7.68    \\
        IQR 2   & 0.39 & 0.66 & 8.03    & 1.93 & 3.68   &  12.03 & 3.11 & 4.35 & 26.20 & 1.88 & 3.14 & 8.76     & 1.16 & 1.98 & 6.23    \\
        \bottomrule
    \end{tabular}
    }
    \label{tab:iqr-l2-ms}
\end{table}

\begin{table}[ht!]
    \centering
    \caption{Comparison of MAE, RMSE, and \%RMSE for $V^{(2)}$ of EFG tensor in the test sets of the initial dataset and those cleaned with IQR with scale factors k = 2, 3, and 4.}
    \resizebox{\textwidth}{!}{
    \begin{tabular}{c c c c c c c c c c c c c c c c}
    \toprule
        \multirow{2}{*}{Dataset} & \multicolumn{3}{c}{$^{1}$H} & \multicolumn{3}{c}{$^{17}$O} & \multicolumn{3}{c}{$^{23}$Na} & \multicolumn{3}{c}{$^{27}$Al} & \multicolumn{3}{c}{$^{29}$Si} \\
        \cmidrule(l{3pt}r{3pt}){2-4}
        \cmidrule(l{3pt}r{3pt}){5-7}
        \cmidrule(l{3pt}r{3pt}){8-10}
        \cmidrule(l{3pt}r{3pt}){11-13}
        \cmidrule(l{3pt}r{3pt}){14-16}
                & {MAE}   & {RMSE} & {\%RMSE}  & {MAE}     & {RMSE}  & {\%RMSE}  & {MAE}   & {RMSE}  & {\%RMSE}  & {MAE}   & {RMSE}  & {\%RMSE} & {MAE}   & {RMSE} & {\%RMSE} \\
        \midrule
        Full    & 0.002 & 0.003 & 1.30  & 0.011 & 0.025 & 4.25  & 0.019 & 0.026 & 19.31 & 0.011 & 0.018 & 4.62  & 0.007 & 0.013 & 3.40 \\
        IQR 4   & 0.002 & 0.003 & 1.27  & 0.009 & 0.022 & 3.71  & 0.018 & 0.025 & 18.68 & 0.009 & 0.019 & 4.81  & 0.006 & 0.011 & 2.98 \\
        IQR 3   & 0.001 & 0.003 & 1.35  & 0.008 & 0.019 & 3.31  & 0.019 & 0.026 & 20.01 & 0.008 & 0.014 & 3.69  & 0.005 & 0.009 & 2.53 \\
        IQR 2   & 0.002 & 0.003 & 1.41  & 0.008 & 0.022 & 3.71  & 0.019 & 0.027 & 20.62 & 0.009 & 0.017 & 4.63  & 0.005 & 0.009 & 2.81 \\
        \bottomrule
    \end{tabular}
    }
    \label{tab:iqr-l2-efg}
\end{table}

\begin{table}[ht!]
    \centering
    \caption{Number of each atom type in the training, validation and test sets for the database used in the final model, i.e. cleaned with IQR with k=2}
    \begin{tabular}{l *{5}{S[table-format=6.0, group-separator={,}]}}
        \toprule
                &   {Hydrogen}   &   {Oxygen}   &   {Sodium}  &   {Aluminium}      &   {Silicon}  \\
        \midrule
        Training   &   50750   &   223568  &   2787    &   15605   &   86696   \\
        Validation   &   6110    &   27557   &   385     &   1881    &   10744   \\
        Test    &   6477    &   28017   &   382     &   1953    &   10829   \\
        Total   &   63337   &   279142  &   3554    &   19439   &   108269  \\
        \bottomrule
    \end{tabular}
    \label{tab:num-atom-types}
\end{table}

\FloatBarrier
\section{Parity plots for train and validation sets}
\begin{figure}[ht!]
  \centering
  \includegraphics[width=\textwidth]{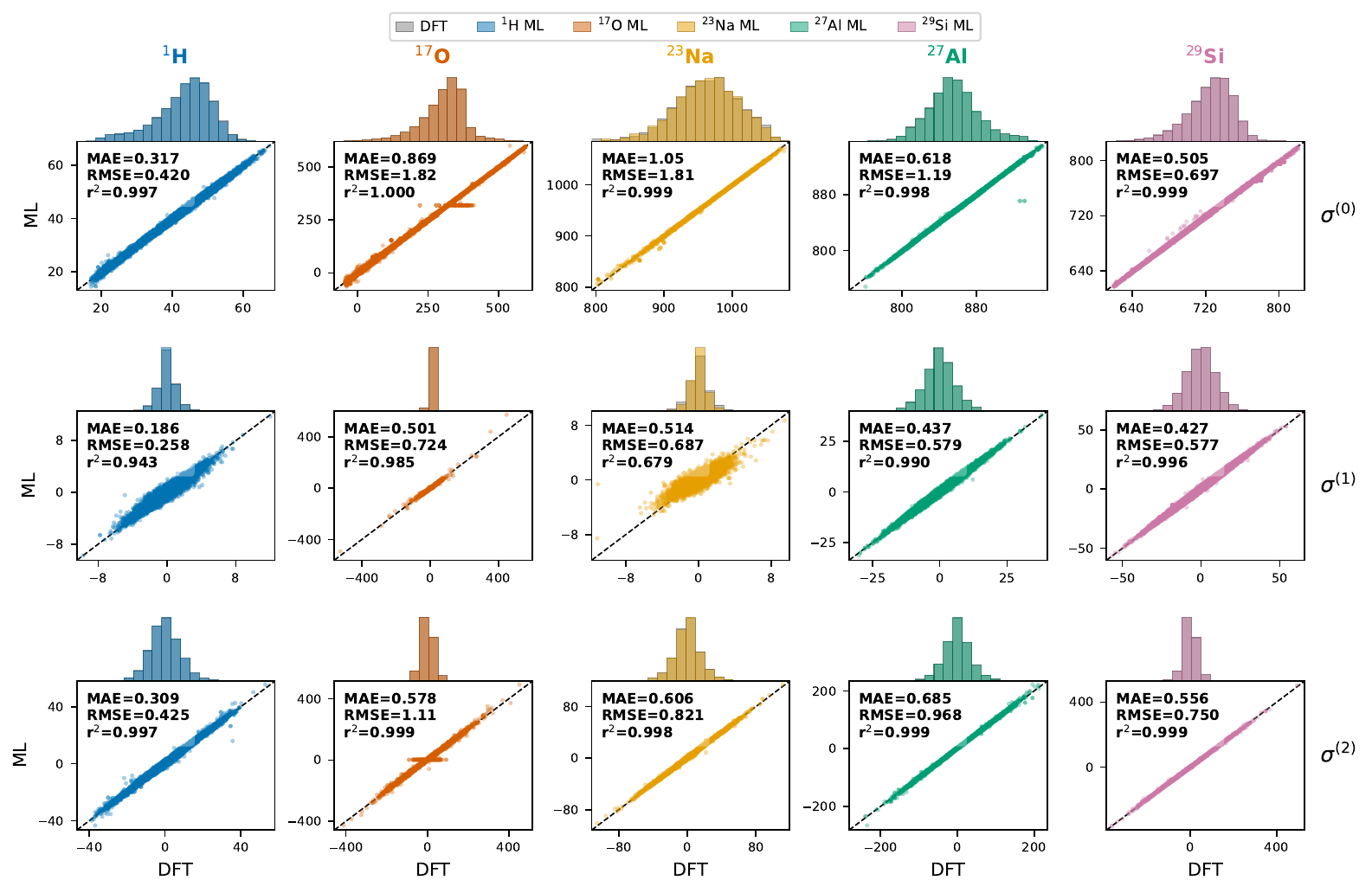}
  \caption{Parity plots for training set comparing DFT-calculated and ML-predicted irreducible spherical tensor components ($\sigma^{(0)}$, $\sigma^{(1)}$, and $\sigma^{(2)}$) of the magnetic shielding tensor for $^{1}$H, $^{17}$O, $^{23}$Na, $^{27}$Al, and $^{29}$Si on the train set. Marginal histograms show the distribution of DFT (gray) and ML (colored) values. All values are in ppm.}
  \label{fig_si:train-parity-ms}
\end{figure}

\begin{figure}[ht!]
  \centering
  \includegraphics[width=\textwidth]{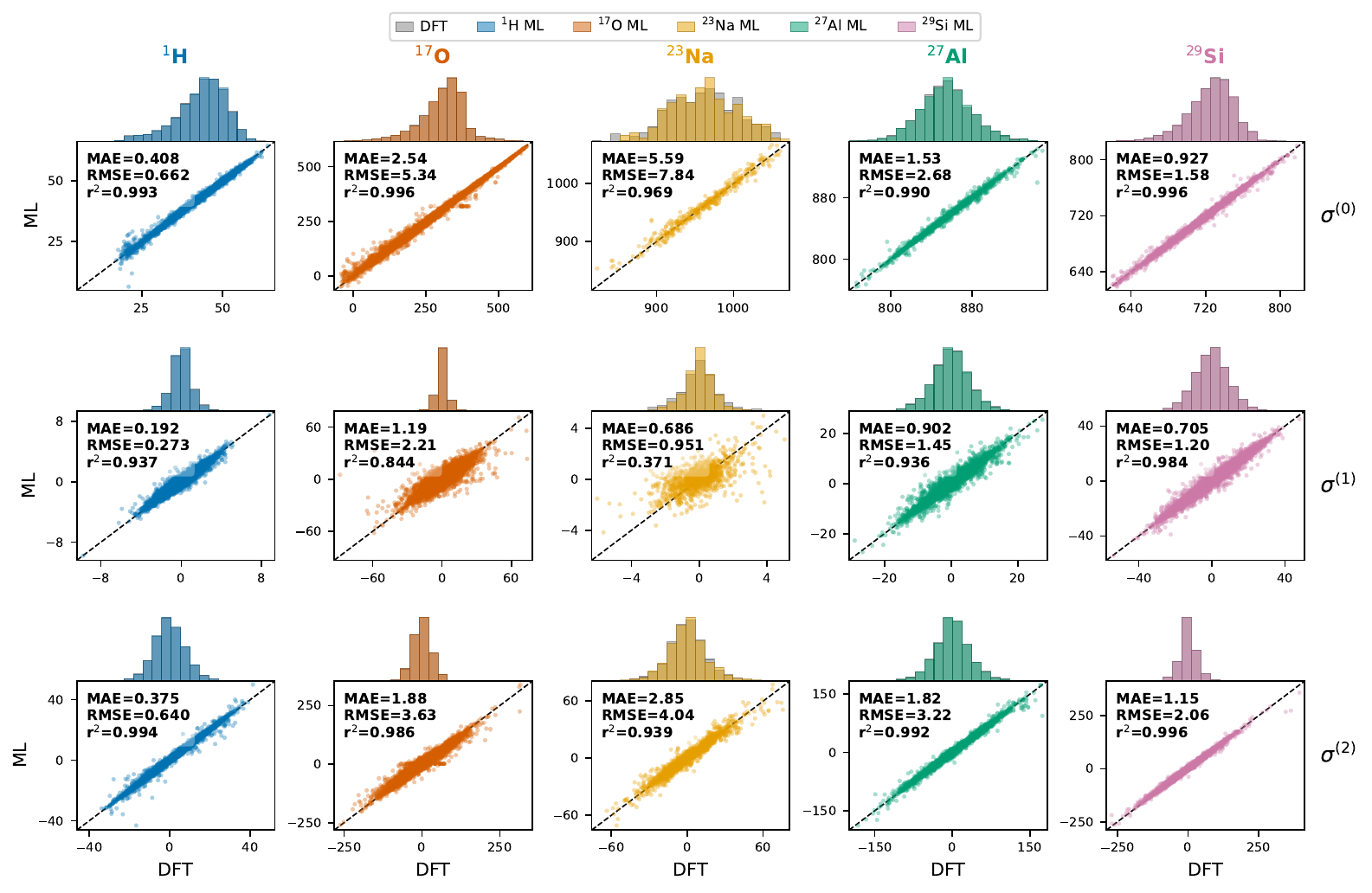}
  \caption{Parity plots for validation set comparing DFT-calculated and ML-predicted irreducible spherical tensor components ($\sigma^{(0)}$, $\sigma^{(1)}$, and $\sigma^{(2)}$) of the magnetic shielding tensor for $^{1}$H, $^{17}$O, $^{23}$Na, $^{27}$Al, and $^{29}$Si on the validation set. Marginal histograms show the distribution of DFT (gray) and ML (colored) values. All values are in ppm.}
  \label{fig_si:valid-parity-ms}
\end{figure}

\begin{figure}[ht!]
  \centering
  \includegraphics[width=\textwidth]{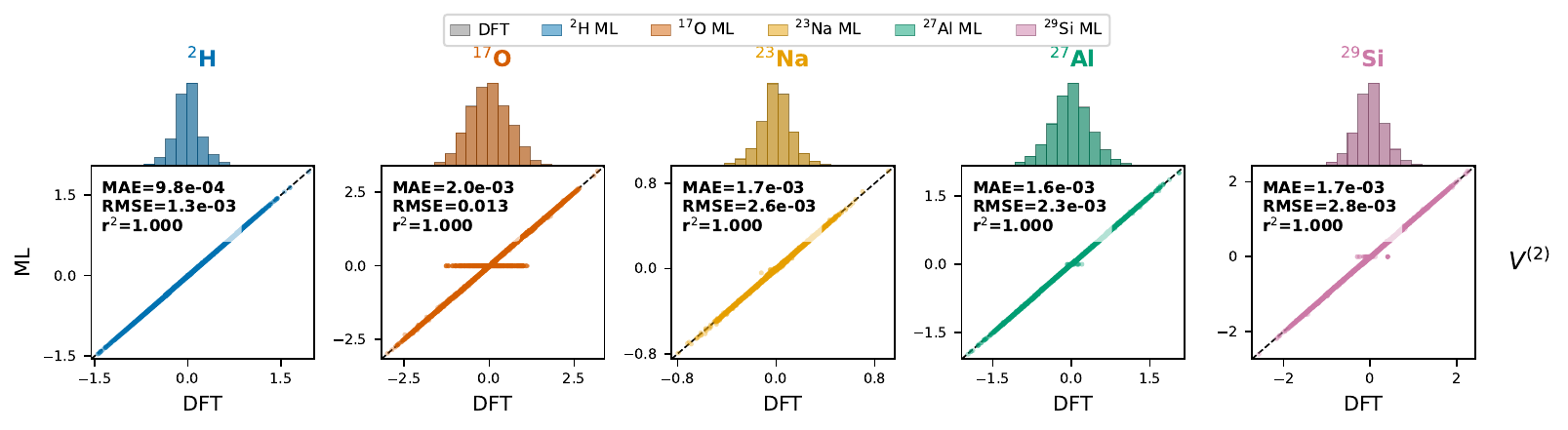}
  \caption{Parity plots for the training set comparing DFT-calculated and ML-predicted $V^{(2)}$ components of the electric field gradient tensor for $^{1}$H, $^{17}$O, $^{23}$Na, $^{27}$Al, and $^{29}$Si. The EFG tensor is symmetric and traceless and therefore contains only $V^{(2)}$ components. Marginal histograms show the distribution of DFT (gray) and ML (colored) values. All values are in atomic units.}
  \label{fig_si:train-parity-efg}
\end{figure}

\begin{figure}[ht!]
  \centering
  \includegraphics[width=\textwidth]{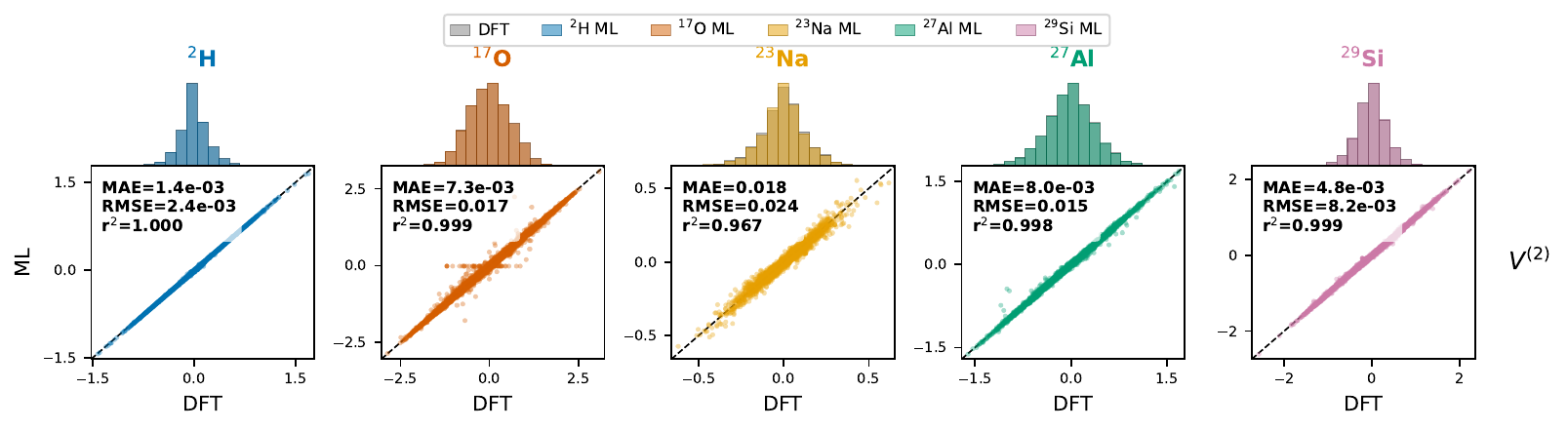}
  \caption{Parity plots for the validation set comparing DFT-calculated and ML-predicted $V^{(2)}$ components of the electric field gradient tensor for $^{1}$H, $^{17}$O, $^{23}$Na, $^{27}$Al, and $^{29}$Si. The EFG tensor is symmetric and traceless and therefore contains only $V^{(2)}$ components. Marginal histograms show the distribution of DFT (gray) and ML (colored) values. All values are in atomic units.}
  \label{fig_si:valid-parity-efg}
\end{figure}

\FloatBarrier
\section{Parity plots for tensor observables (train and validation sets)}

\begin{figure}[ht!]
  \centering
  \includegraphics[width=\textwidth]{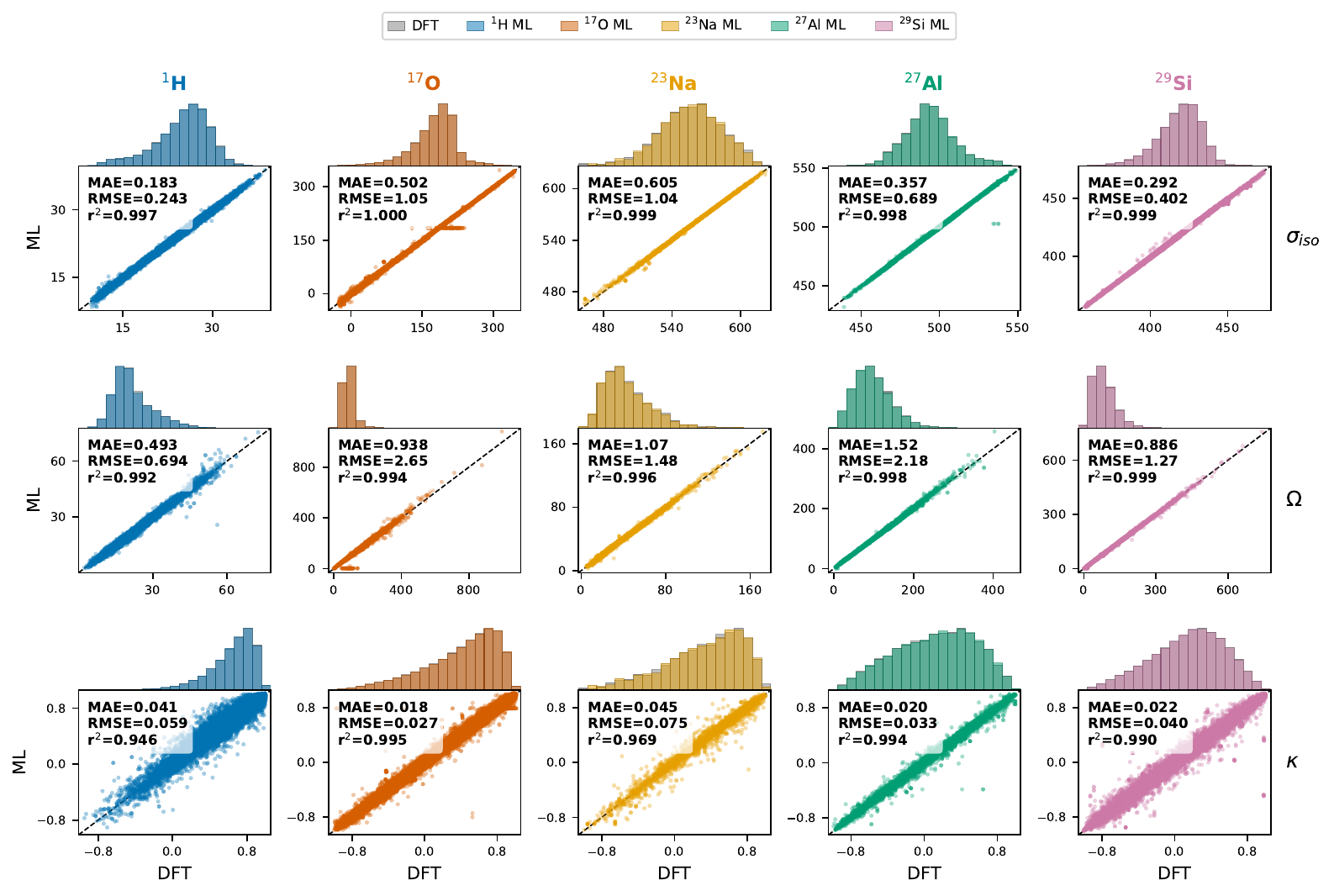}
  \caption{Parity plots for the training set comparing DFT-calculated and ML-predicted magnetic shielding tensor observables ($\sigma_\mathrm{iso}$, $\Omega$, and $\kappa$) for $^{1}$H, $^{17}$O, $^{23}$Na, $^{27}$Al, and $^{29}$Si. All values are in ppm.}
  \label{fig_si:train-parity-ms-observables}
\end{figure}

\begin{figure}[ht!]
  \centering
  \includegraphics[width=\textwidth]{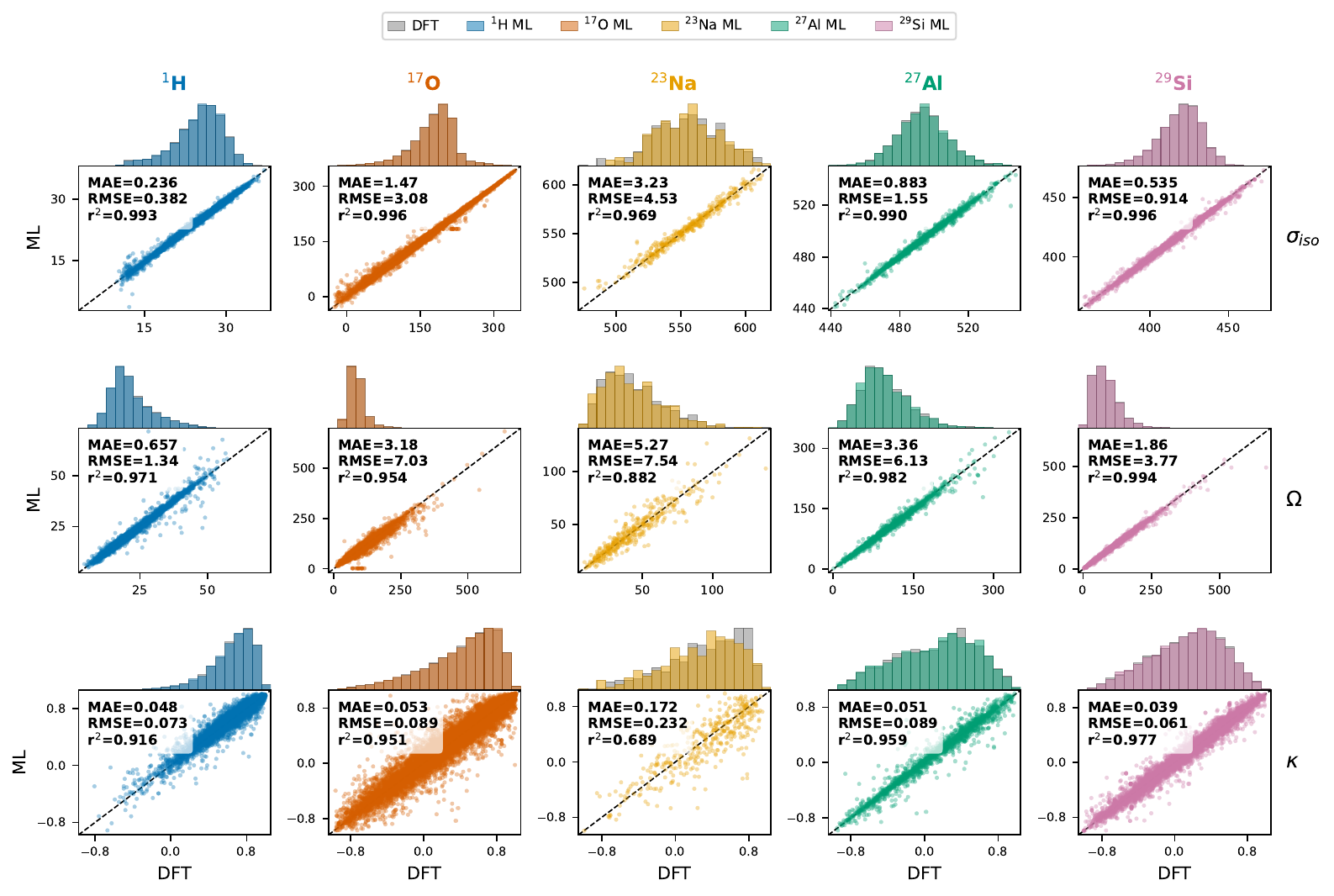}
  \caption{Parity plots for the validation set comparing DFT-calculated and ML-predicted magnetic shielding tensor observables ($\sigma_\mathrm{iso}$, $\Omega$, and $\kappa$) for $^{1}$H, $^{17}$O, $^{23}$Na, $^{27}$Al, and $^{29}$Si. All values are in ppm.}
  \label{fig_si:valid-parity-ms-observables}
\end{figure}

\begin{figure}[ht!]
  \centering
  \includegraphics[width=\textwidth]{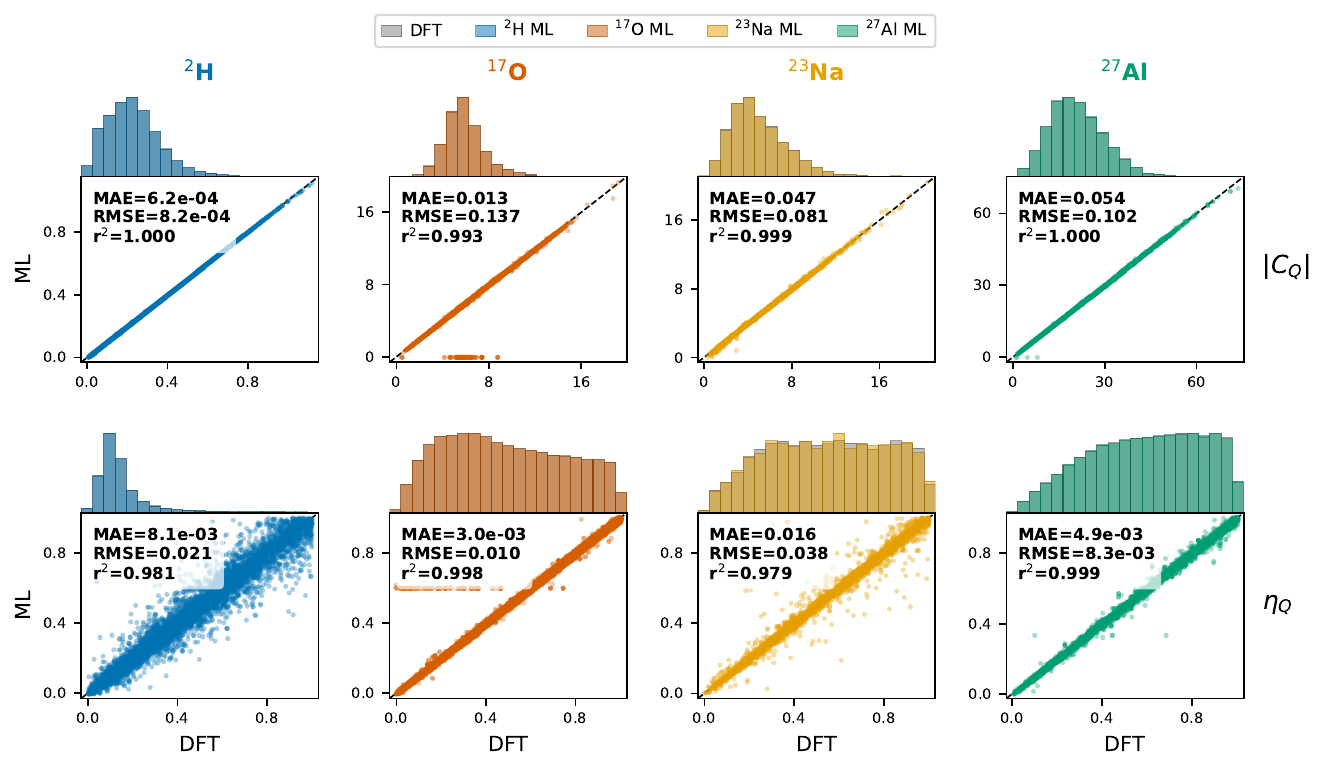}
  \caption{Parity plots for the training set comparing DFT-calculated and ML-predicted EFG tensor observables ($|C_Q|$ and $\eta_Q$) for $^{2}$H, $^{17}$O, $^{23}$Na, and $^{27}$Al. $C_Q$ values are in MHz; $\eta_Q$ is dimensionless.}
  \label{fig_si:train-parity-efg-observables}
\end{figure}

\begin{figure}[ht!]
  \centering
  \includegraphics[width=\textwidth]{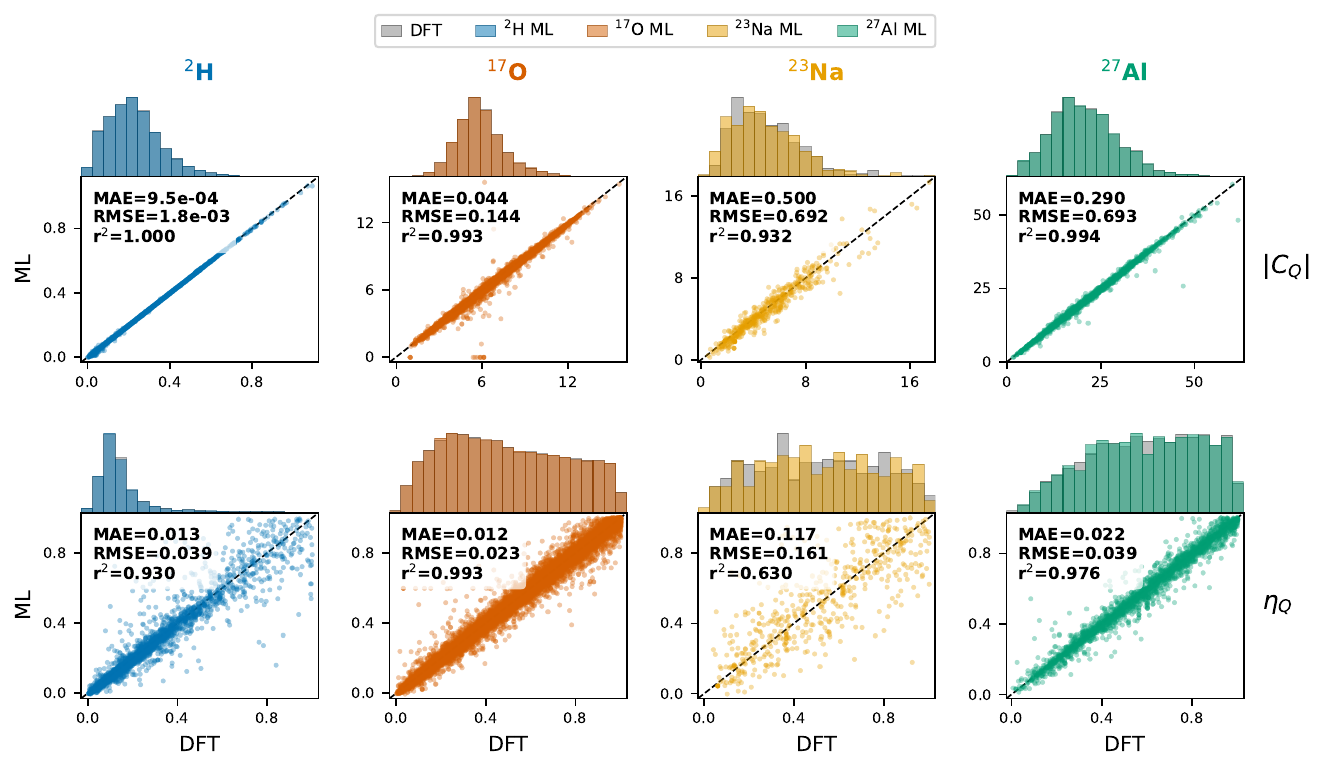}
  \caption{Parity plots for the validation set comparing DFT-calculated and ML-predicted EFG tensor observables ($|C_Q|$ and $\eta_Q$) for $^{2}$H, $^{17}$O, $^{23}$Na, and $^{27}$Al. $C_Q$ values are in MHz; $\eta_Q$ is dimensionless.}
  \label{fig_si:valid-parity-efg-observables}
\end{figure}

\begin{figure}[ht!]
  \centering
  \includegraphics[width=\textwidth]{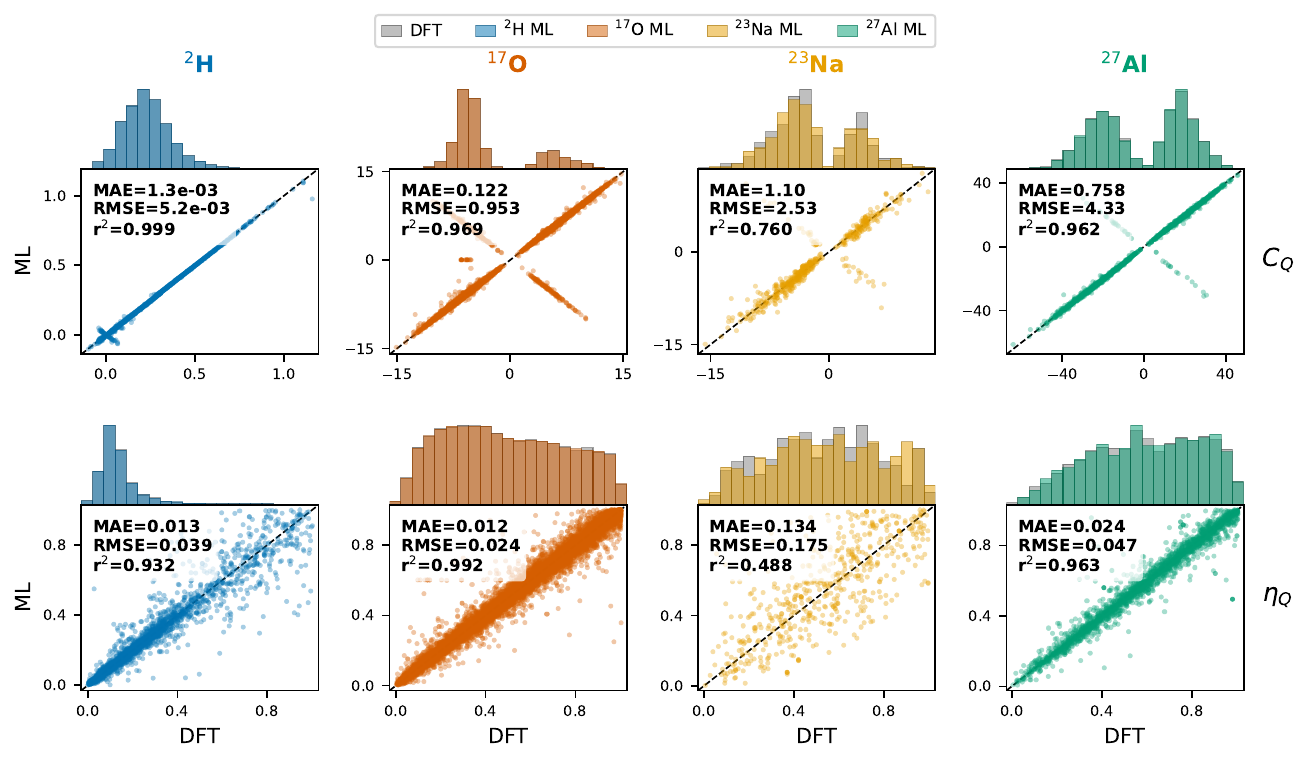}
  \caption{Parity plots comparing DFT-calculated and ML-predicted signed $C_Q$ and $\eta_Q$ for $^{2}$H, $^{17}$O, $^{23}$Na, and $^{27}$Al on the test set. Unlike Figure~5 in the main text, $C_Q$ is shown without taking the absolute value. The cross-shaped scatter patterns visible for $^{17}$O and especially $^{23}$Na indicate instances where the model predicts the wrong sign of $C_Q$. This sign ambiguity is experimentally irrelevant, as standard solid-state NMR spectra are invariant to the sign of $C_Q$.}
  \label{fig_si:test-parity-efg-observables}
\end{figure}

\FloatBarrier
\begin{table}[ht!]
    \centering
    \caption{Comparison between DFT-calculated and ML-predicted $^{27}$Al isotropic chemical shifts ($\delta_\mathrm{iso}$) and quadrupolar coupling constants ($C_Q$) for dehydrated Al-containing H-RTH. Values are time-averaged from 1~ns ML-MD trajectories. No experimental data are available for direct comparison.}
    \begin{tabular}{c c c c c}
        \toprule
        \multirow{2}{*}{T-site}  & \multicolumn{2}{c}{DFT}     & \multicolumn{2}{c}{ML} \\
        \cmidrule(l{3pt}r{3pt}){2-3}
        \cmidrule(l{3pt}r{3pt}l{3pt}r{3pt}){4-5}
                & $\delta_{iso}$ (ppm)    & $C_Q$ (MHz) & $\delta_{iso}$ (ppm)    & $C_Q$ (MHz) \\
        \midrule
        T1  & 60.05 & 17.26 & 59.43 & 17.36 \\
        T2  & 59.23	& 18.12	& 59.04	& 18.20	\\
        T3	& 61.23	& 17.11	& 60.90	& 17.16  \\
        T4	& 61.47	& 17.33	& 61.35	& 17.22  \\
        \bottomrule
    \end{tabular}
    \label{tab:rth-al-dehyd}
\end{table}

\FloatBarrier